\documentclass[fleqn,usenatbib]{mnras}

\usepackage{newtxtext,newtxmath}

\usepackage[T1]{fontenc}
\usepackage{ae,aecompl}

\usepackage{etoolbox}
\makeatletter
\patchcmd\@combinedblfloats{\box\@outputbox}{\unvbox\@outputbox}{}{\errmessage{\noexpand patch failed}}
\makeatother

\usepackage{graphicx}	
\usepackage{amsmath}	
\DeclareMathOperator{\sech}{sech}
\usepackage{amssymb}	
\usepackage[figure,figure*]{hypcap}
\usepackage{float}

\newcommand{\comm}[1]{}
\defcitealias{2019MNRAS.490.1539R}{R19}

\title[Tilted structures in edge-on galaxies?]{Tilted outer and inner structures in edge-on galaxies?}

\author[A. V. Mosenkov et al.]{
Aleksandr V. Mosenkov,$^{1,2}$\thanks{E-mail: aleksandr.mosenkov@tdtu.edu.vn}
Anton A. Smirnov,$^{3,4}$
Olga K. Sil'chenko,$^{5}$
\newauthor
R. Michael Rich,$^{6}$
Vladimir P. Reshetnikov,$^{3,7}$
and
John Kormendy$^8$
\\
$^{1}$Informetrics Research Group, Ton Duc Thang University, Ho Chi Minh City, Vietnam\\
$^{2}$Faculty of Applied Sciences, Ton Duc Thang University, Ho Chi Minh City, Vietnam\\
$^{3}$St.Petersburg State University, 7/9 Universitetskaya nab., St.Petersburg, 199034, Russia\\
$^{4}$Central Astronomical Observatory at Pulkovo of RAS, Pulkovskoye Chaussee 65/1, 196140 St. Petersburg, Russia\\
$^{5}$Sternberg Astronomical Institute of the Lomonosov Moscow State University, University av. 13, Moscow, 119234, Russia\\
$^{6}$Department of Physics \& Astronomy, Univ. of California Los Angeles, 430 Portola Plaza, Los Angeles, CA 90095-1547, USA\\
$^{7}$Special Astrophysical Observatory, Russian Academy of Sciences, 369167 Nizhnij Arkhyz, Russia\\
$^{8}$Department of Astronomy, University of Texas at Austin, 2515 Speedway, Mail Stop C1400, Austin, TX 78712, USA\\\\
}

\date{Accepted XXX. Received YYY; in original form ZZZ}

\pubyear{2020}

\begin{document}
\label{firstpage}
\pagerange{\pageref{firstpage}--\pageref{lastpage}}
\maketitle

\begin{abstract}
Tilted and warped discs inside tilted dark matter haloes are predicted from numerical and semi-analytical studies. In this paper, we use deep imaging to demonstrate the likely existence of tilted outer structures in real galaxies. We consider two SB0 edge-on galaxies, NGC\,4469 and NGC\,4452, which exhibit apparent tilted outer discs with respect to the inner structure. In NGC\,4469, this structure has a boxy shape, inclined by $\Delta \mathrm{PA}\approx3\degr$ with respect to the inner disc, whereas NGC\,4452 harbours a discy outer structure with $\Delta \mathrm{PA}\approx6\degr$. In spite of the different shapes, both structures have surface brightness profiles close to exponential and make a large contribution ($\sim30$\%) to the total galaxy luminosity. In the case of NGC\,4452, we propose that its tilted disc likely originates from a former fast tidal encounter (probably with IC\,3381). For NGC\,4469, a plausible explanation may also be galaxy harassment, which resulted in a tilted or even a tumbling dark matter halo. A less likely possibility is accretion of gas-rich satellites several Gyr ago. New deep observations may potentially reveal more such galaxies with tilted outer structures, especially in clusters. We also consider galaxies, mentioned in the literature, where a central component (a bar or a bulge) is tilted with respect to the stellar disc. According to our numerical simulations, one of the plausible explanations of such observed ``tilts'' of the bulge/bar is a projection effect due to a not exactly edge-on orientation of the galaxy coupled with a skew angle of the triaxial bulge/bar. 
\end{abstract}




\section{Introduction}
\label{sec:intro}
The structure of disc galaxies evolves by various internal \citep{2004ARA&A..42..603K,2013seg..book....1K} and external  \citep{2012ApJS..198....2K} processes. Galaxy discs can be flared, warped, tilted and truncated \citep{2011ARA&A..49..301V}. All these prominent structural features can be explained by various evolution mechanisms, no one of which is universally dominant.

Galactic warps, seen as large-scale distortions of stellar and gaseous discs, when, starting at some radius, the tips of the outer isophotes start to bend, are very common in the Local Universe \citep{1998A&A...337....9R} and were very frequent in the past \citep{2002A&A...382..513R}. Optical observations show that at least 40-50\% of disc galaxies exhibit warps to some degree \citep{1990MNRAS.246..458S,1998A&A...337....9R,2006NewA...11..293A,2016JKAS...49..239A}. Galactic warps are even more conspicuous in H{\sc i} \citep{1981AJ.....86.1825B,1990ApJ...352...15B} and seen by H{\sc ii} regions \citep{2004MNRAS.347..237P,2014A&A...569A.125H}. The main proposed mechanisms to explain this phenomenon are misalignment between the disc and dark halo \citep{1999ApJ...513L.107D,2000MNRAS.311..733I}, accretion onto the disc \citep{1989MNRAS.237..785O,1999MNRAS.303L...7J,2006MNRAS.370....2S,2010MNRAS.408..783R}, tidal interaction due to fast encounters \citep{2000ApJ...534..598V,2012ApJS..198....2K,2014ApJ...789...90K,2020arXiv200207022S}, intergalactic magnetic fields \citep{1990A&A...236....1B,1998A&A...332..809B,2010A&A...519A..53G}, and discrete bending modes \citep{1988MNRAS.234..873S,2004A&A...425...67R}. 

Another important process, disc tilting, is a change of the overall angular momentum (vector) of the disc with time whereas disc warping implies that the direction of the angular momentum vector changes with galactocentric radius. In essence, disc tilting manifests itself as a slewing of the disc plane with time, with respect to its current orientation in the space. There are different processes which can cause galactic discs to tilt \citep[see e.g. introduction in][]{2019MNRAS.488.5728E}: minor merging events \citep{1975ApJ...202L.113O,1997ApJ...480..503H,2009ApJ...700.1896K,2012MNRAS.420.3324B}, tumbling of dark matter haloes \citep{2004ApJ...616...27B,2007MNRAS.380..657B,2012MNRAS.426..983D}, and gas cooling onto the disc plane \citep{2015MNRAS.452.4094D,2019MNRAS.488.5728E}. If a galaxy experiences an infall of a satellite, its disc also exhibits two other dynamical responses: warping and thickening \citep{1997ApJ...480..503H,2008MNRAS.389.1041R,2009ApJ...700.1896K,2014MNRAS.442..160S}. Tilting cannot be observed directly due to its extremely low tilting rate, even as compared to a typical galaxy rotation period ($\sim 5\degr$\,Gyr$^{-1}$, \citealt{2004ApJ...616...27B,2015MNRAS.452.2367Y}). However, the existence of some distinctive features in the disc structure, such as conspicuous S-shape warps or strong disc thickening may indirectly point to disc tilting or the result of a satellite accretion. 

It was found that the angular momentum of the hot gas corona around galaxies, shock heated due to the falling of external gas into the dark matter's potential well, is usually misaligned with that of their stellar disc \citep{2002ApJ...576...21V,2010MNRAS.408..783R,2015MNRAS.453..721V,2017MNRAS.467.2066S}. When the hot coronal gas cools, it settles into the disc \citep{1980MNRAS.193..189F,2005MNRAS.363....2K} and contributes misaligned angular momentum to the disc, which results in the disc tilting \citep{2015MNRAS.452.4094D,2019MNRAS.488.5728E}. Consequently, it was found that the principal axes of dark matter haloes in blue star-forming galaxies are generally misaligned with the principal axes of their discs \citep{2006MNRAS.369.1293Y,2008MNRAS.385.1511W}.

According to kinematic studies of nearby lenticular galaxies, misalignments between their stellar discs and gaseous components (from 40\% to 60\% of early-type galaxies have cool and ionized gas, \citealt{2003ApJ...584..260W,2006ApJ...644..850S,2010ApJ...725..100W,2011MNRAS.417..882D,2012MNRAS.422.1835S}) are frequent \citep[see e.g.][]{1992ApJ...401L..79B,1996MNRAS.283..543K,2006AJ....131.1336S,2009ApJ...694.1550S,2011ApJ...740...83K,2011MNRAS.417..882D,2013ApJ...769..105K,2015AJ....150...24K,2020A&A...634A.102P}. \citet{2011MNRAS.417..882D} showed that misaligned gaseous subsystems are four times more frequent in field lenticular galaxies than in those residing in a cluster. Among strictly isolated nearby S0s, \citet{2014MNRAS.438.2798K} found that half of all ionized-gas discs counterrotates the stars \citep[see also][]{2015AJ....150...24K}. Recently, \citet{2019ApJS..244....6S} studied a sample of 18 S0 galaxies and found five galaxies with strongly inclined ionized gas discs.
Some fraction of the gas observed in many S0s seems to be accreted in recent events, either due to tidal disruptions of massive gas-rich satellites \citep{2009MNRAS.394.1713K,2011MNRAS.411.2148K} or gas accretion from cosmological filaments \citep{1996ApJ...461...55T,1998ApJ...506...93T,2005MNRAS.363....2K,2006MNRAS.368....2D}.

\begin{figure}
\label{fig:sketch}
\centering
\includegraphics[width=8cm]{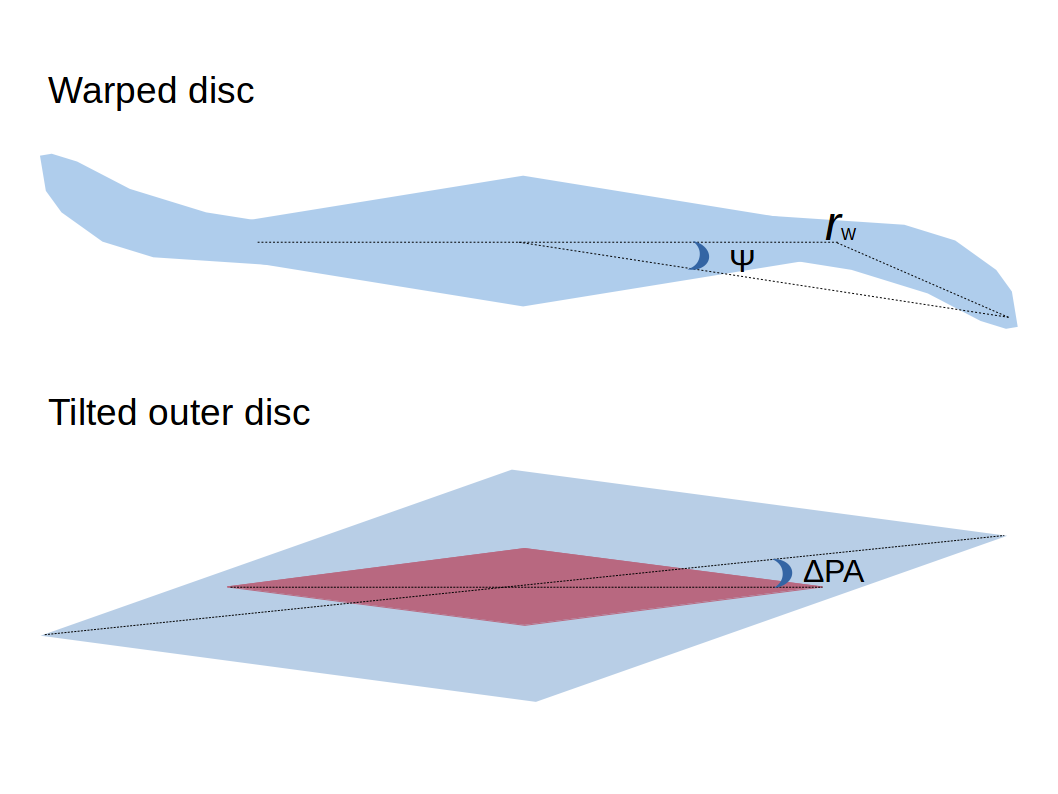}
\caption{Difference between disc tilting and warping. The warp angle $\Psi$ and the warp radius $r_\mathrm{W}$, where a warp starts, are also shown. The angle between the two different planes of the inner and outer discs in the bottom sketch is shown as $\Delta$PA.}
\end{figure}

Another interesting effect of disc tilting might be misalignment between inner and outer disc structures in edge-on galaxies.
In this paper by inner and outer structures we mean that they are separated not only radially, but also vertically, so that an inner structure is enclosed in an outer structure. By tilting of a component we will call an observational effect where this component, as a whole, shows a different position angle than the general galaxy plane. Note here that we distinguish tilting from warping, which manifests itself in a vertical distortion of the disc starting at some radius, though both can have the same origin. The difference between disc warps and tilts is illustrated in Fig.~\ref{fig:sketch}. To this day, based on deep imaging, there has been no convincing observational evidence that the rather faint outer \textit{stellar} structure in edge-on galaxies may exhibit tilting relative to the inner region in the sense that the inner and outer structures lie in different planes (see, however, our note below). Edge-on galaxies, as the best targets to study the vertical galaxy structure and disc tilting, were extensively studied in the optical and near-infrared \citep{1996A&AS..117...19D,2002MNRAS.334..646K,2010MNRAS.401..559M,2011ApJ...741...28C, 2012MNRAS.427.1102M,2014ApJ...787...24B,2016MNRAS.459.1276C,2018A&A...610A...5C}, including deep observations \citep{2010AJ....140..962M,2011A&A...536A..66M,2018A&A...614A.143M,2019ApJ...883L..32V,2019arXiv191005358G}, but no systematic difference between the position angles in the inner and faint outer galaxy regions was noticed, except for \citet{ 2020MNRAS.494.1751M} where they studied the outer shape of edge-on galaxies and found several galaxies with tilted outer isophotes. The current paper is a follow-up study of \citet{2020MNRAS.494.1751M}. 

\begin{figure*}
\label{fig:NGC3034_image}
\centering
\includegraphics[width=8cm]{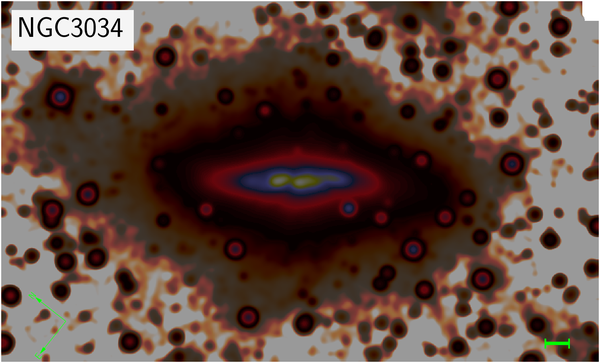}
\includegraphics[width=8cm]{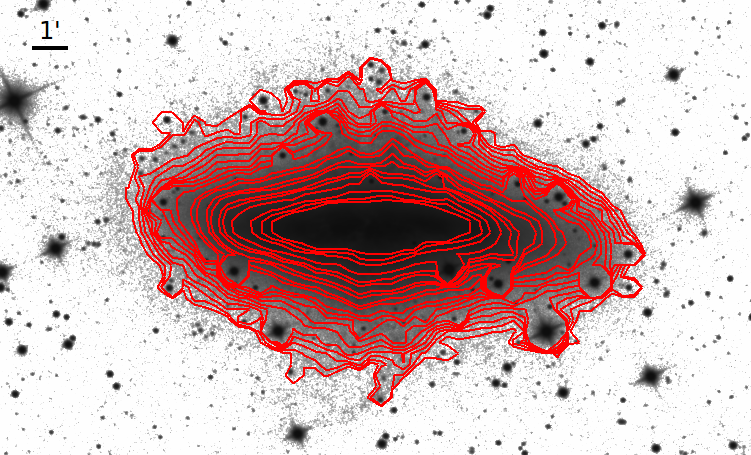}
\caption{HERON image for NGC\,3034 (left plot) and its superimposed isophotes from 20 to 26 mag/arcsec$^2$ (right plot).}
\end{figure*}

In the present paper, we consider a surface photometry of two SB0 galaxies with an apparent tilt of the outer structure with respect to the inner one as it represents a different disc which has a different position angle than the inner disc. 
NGC\,4469 has been noted in \citet{2020MNRAS.494.1751M}, where they investigated deep imaging of 35 edge-on galaxies selected from the Halos and Environments of Nearby Galaxies (HERON) survey \citep{2019MNRAS.490.1539R}. We should note that M\,82, the Cigar galaxy, which was also observed in the framework of the HERON survey, likewise shows tilted isophotes in the periphery (see Fig.~\ref{fig:NGC3034_image}), which may be associated with its spiral arms \citep{2005ApJ...628L..33M} or a disc warp \citep{1983IAUS..100...67N}. Also, this galaxy shows evident signs of interaction with its neighbour M\,81 \citep{1993ApJ...411L..17Y}. Taking into account the foregoing arguments and the peculiar nature of this well-known starburst galaxy \citep{1980ApJ...238...24R,1990ApJS...74..833H}, we decided not to consider M\,82 in our study. Another galaxy with a prominent tilted outer structure, NGC\,509, which was classified by \citet{2020MNRAS.494.1751M} as an edge-on galaxy, is in fact not so highly inclined, therefore we rejected it as well. In this study we do not consider another remarkable edge-on galaxy from the HERON survey with an apparent tilt of the outer structure, NGC\,4638, which exhibits a tilt of the bright stellar halo with respect to the disc. We are about to consider this galaxy among other edge-on galaxies with tilted bright haloes in our further paper, whereas in this study we mainly focus on tilted disc components.
In addition to NGC\,4469, NGC\,4452 was chosen for this study incidentally, during a visual inspection of the EGIS catalogue of edge-on galaxies \citep{2014ApJ...787...24B}. Using the technique described in Sect.~\ref{sec:data} (see also \citealt{2018A&A...614A.143M}), we are likely to find more objects with such faint outer structures in our further studies. 

\citet{1982SAAS...12..115K} discussed isophote twists in non-edge-on galaxies which can be the result of tidal effects due to a close (or close in the past) companion. The best example is the spheroidal galaxy NGC\,205 which exhibits an isophote twist toward M\,31. \citet{1993A&AS...98...29M} (see also \citealt{1994A&AS..105..481M}) found small twists of the isophotes for 10 nearly edge-on S0 galaxies, which are manifested in small trends in their major axis position angles and one of the Fourier harmonics responsible for the isophote asymmetry. They attributed this change of the position angle to a small difference of the major axis orientation between the central component (a bulge and/or bar) and an outer component (a disc or a halo). They estimate this tilt of a few degrees between the equatorial planes of these components, and claim that it is not associated with spiral arms and disc warps. Here we should note, however, that we revisited a photometry of these galaxies using the data preparation method described in Sect.~\ref{sec:data} and established that all galaxies, except for NGC\,4638, show either an inner twist of an inner triaxial structure (possibly, a bar) or `a tilt' of the halo, but the estimated edge-on orientation of these galaxies is either questionable (e.g. NGC\,2732 and NGC\,7332) or not true (e.g. NGC\,4036 and NGC\,4251).  It is evident that when there are triaxial components in a flat disc that has one tilt angle to the line of sight, then isophotes at different surface brightnesses will necessarily have different position angles. But how important this effect is when we consider almost edge-on galaxies? Do we observe edge-on galaxies with inner components with position angles that are different from that of the main disc? For a pedagogical purpose, in this paper we consider several not exactly edge-on galaxies (including NGC\,509 in Appendix~\ref{Appendix:NGC509}) which exhibit tilting of the inner structure similar to those reported by \citet{1993A&AS...98...29M}. We show that these structures may be, in fact, bars which appear tilted due to the projection effect.

This paper is organized as follows. In Sect.~\ref{sec:data} we describe the data and image preparation for the two selected galaxies with possible titled outer discs. In Sect.~\ref{galaxies}, we consider in detail each of the selected galaxies. We discuss our results in Sect.~\ref{sec:discussion} and make final conclusions in Sect.~\ref{sec:conclusions}.

\section{The data}
\label{sec:data}

Here we describe the data for NGC\,4452 and NGC\,4469, which we use in our photometric analysis in Sect.~\ref{galaxies}. 
For NGC\,4469, its deep image has been considered in the framework of the HERON survey (\citealt{2019MNRAS.490.1539R, 2020MNRAS.494.1751M}). However, here we analyse observations of both galaxies obtained from the Sloan Digital Sky Survey DR16 \citep[SDSS, ][]{2000AJ....120.1579Y,2019arXiv191202905A} and the DESI Legacy Imaging Surveys DR8 (hereafter Legacy, \citealt{2019AJ....157..168D}). This allows us to 1) explore the faint outer structure of these galaxies down to the same level of depth, and 2) study their colour maps. Also, the resolution in these surveys is better than in the HERON survey, although their imaging is shallower (below, however, we describe how we increase their depth by stacking galaxy images in several optical wavebands, as proposed in \citealt{2011A&A...536A..66M} and \citealt{2018A&A...614A.143M}). In principal, we could limit ourselves to use SDSS but Legacy has deeper imaging and a better resolution (see below), therefore we decided to use both data sources. Our image preparation is organized as follows\footnote{It is realised with a dedicated Python package {\tt{IMAN}} \url{https://bitbucket.org/mosenkov/iman\_new}}.

Using a special Python script to download and concatenate adjacent fields from the SDSS archive\footnote{\url{https://github.com/latrop/sdss_downloader}}, we retrieve galaxy images for both galaxies in the $gri$ bands (the $u$ and $z$ bands are rather shallow, \citealt{1996AJ....111.1748F}, so we do not use them) in an automated regime. Also, we download corresponding Point Spread Function (PSF) images (extracted from the respective \textit{psFields} files) in the same wavebands. These tiny images describe the core PSF (seeing), but for the wings of the scattered light we use an extended PSF from \citet{2020MNRAS.491.5317I}, which we rotate to align the drift scanning direction of the image with the drift scanning direction of the PSF. We then merge the inner (core) and outer (extended) PSFs by normalising them in an annulus $5\arcsec$ in width where the two PSFs overlap. After that, the galaxy images in the $g$ and $i$ bands are resampled to the $r$ band using the created extended PSFs by means of the {\tt{pypher}}\footnote{\url{https://github.com/aboucaud/pypher}} package \citep{2016A&A...596A..63B} suited for PSF matching.  

After that, for each galaxy image we create a segmentation map using {\tt{SExtractor}} \citep{1996A&AS..117..393B} and increase the size of the regions of the general mask by a factor of two, to ensure that our mask covers all faint scattered light in the image. Then we apply the created mask and fit the background with a first order polynomial and subtract it from the initial image. Finally, we stack the galaxy images in all three bands --- this increases the signal-to-noise ratio and, consequently, reduces the root-mean-square (rms) of the background. For this purpose we use the {\tt{IRAF}}/{\tt{IMCOMBINE}} procedure. In the resultant image we select good PSF (unsaturated) stars and measure the total fluxes of these stars using a procedure from the Python package {\tt{photutils}}. Then we cross-correlate the selected stars with the SDSS database to retrieve their apparent magnitudes in the $r$ band. By so doing, we are able to perform photometric calibration of the stacked image and estimate its zero-point in the $r$ band. The procedure of stacking the three wavebands together allows us to improve the depth of the imaging by 1.1~mag, on average: $26.5$~mag/arcsec$^2$ for the individual bands versus $27.6$~mag/arcsec$^2$ for the stacked images, calculated at the $3\sigma$ level in a box of $10\times10$~arcsec$^2$ (we make use of this notation from \citealt{2016MNRAS.456.1359F} throughout the paper).

In our final step, we mask out all objects which do not belong to the target galaxy. To do this, we use a special Python script from the {\tt{IMAN}} package which is based on the {\tt{mtobjects}} tool\footnote{\url{https://github.com/CarolineHaigh/mtobjects}} \citep{teeninga2015improved}: it effectively identifies all objects in an image and is robust for masking even in an automatic mode without a special tuning of input parameters.

The same pipeline is applied to the Legacy images in the $grz$ bands, except that we do not do image resampling, as, at the moment, no extended PSFs have been produced for the Legacy imaging. We should note, however, that the image resampling is not a necessary step for enhancing the depth of the images and we only do this to be able to carry out photometric decomposition of the stacked SDSS galaxy images (see Sect.~\ref{galaxies}) using the matched PSFs. The image stacking of the Legacy data allows us to create final images with a depth of $28.3$~mag/arcsec$^2$ calibrated to the $r$ band. 

We also use observations in the NUV band from the
Galaxy Evolution Explorer \citep[GALEX,][]{2005ApJ...619L...1M,2014AdSpR..53..900B} and observations in the W1 band (3.4$\mu$m) from the the Wide-field Infrared Survey Explorer \citep[WISE,][]{2010AJ....140.1868W} to estimate the NUV$-r$ colour and the total stellar mass of the galaxies, respectively (see Sect.~\ref{info}). The images were prepared using the above described routines except for image stacking.

\section{Analysis of the data}
\label{galaxies}

\subsection{The methods}
\label{methods}

For each galaxy, we carry out an {\tt{IRAF}}/{\tt{ELLIPSE}} \citep{1987MNRAS.226..747J} fit of the galaxy isophotes. We consider the distributions of the position angle PA, ellipticity $\epsilon=1-b/a$ (where $a$ and $b$ are the semi-major and semi-minor axes, respectively) and the $B_4$ coefficient, which characterises the shape of the isophotes, see below. Also, we provide a 2D photometric model, consisting of several structural components, using the {\tt{IMFIT}} code \citep{2015ApJ...799..226E}. We use its standard Levenberg-Marquardt algorithm for finding the optimal fit parameters. To estimate indicative uncertainties on our fit parameters, we vary the sky level using a set of Monte-Carlo simulations (10 attempts) with the sky rms determined in Sect.~\ref{sec:data}. For uniformity, to fit the 2D intensity distribution for each structural component of the galaxies under study, we use a S\'ersic function  \citep{1963BAAA....6...41S,1968adga.book.....S} with the following major-axis intensity profile:

\begin{equation}
\label{sersic}
I(r) \; = \; I_\mathrm{e} \: \exp \left\{ -b_\mathrm{n} \left[ \left( \frac{r}{r_\mathrm{e}} \right)^{1/n} \! - \: 1 \right] \right\}\,,
\end{equation}
where $I_\mathrm{e}$ is the surface brightness at the effective (half-light) radius $r_\mathrm{e}$
and $n$ is the S\'ersic index controlling the shape of the intensity profile. The
function $b_\mathrm{n}$ is calculated via the polynomial approximation in
\citet{1999A&A...352..447C} (when $n > 0.36$) and  \citet{2003ApJ...582..689M} (when $n \leq 0.36$).

In our fits we consider generalised ellipses \citep{1990MNRAS.245..130A,2015ApJ...799..226E} with the following free parameters: position angle PA, ellipticity $\epsilon$, and  $C_0$, which controls the discy/boxy view of the isophote (when $C_0<0$ the isophotes look discy, whereas they become boxy if $C_0>0$ and $C_0=0$ if the isophotes can be represented by pure ellipses). Note that one of the output Fourier coefficients of the {\tt{IRAF}}/{\tt{ELLIPSE}} procedure, the $B_4$ coefficient, also characterises the discyness/boxyness of the isophotes, so that when $B_4>0$ the isophotes are discy and they look boxy if $B_4<0$. Apparently, there is a link between $C_0$ and $B_4$, but as they are derived using different approaches, the relation between them is not direct. Thus, we use both when needed.

We denote by $f$ the fraction of light in a given component according to our models.

We should note here that we neglect the influence of dust on the estimated parameters of the photometric components. Although this assumption can be rather strong (as we can see below, NGC\,4469 shows some traces of dust in the central galaxy region), the fraction of bolometric luminosity absorbed by dust in early-type (Hubble stage $T<0.5$) galaxies is estimated to be $7.4\pm0.8$\%, on average, versus $24.9\pm0.7$\% for late-type ($T\geq0.5$) galaxies \citep{2018A&A...620A.112B}. \citet{2012ApJ...748..123S} also show that the stellar discs in S0s contain much less dust than the discs in late-type spirals.

\subsection{General information on the galaxies}
\label{info}

In Table~\ref{tab:main.tab}, we list some general characteristics for both galaxies using their stacked images calibrated with the corresponding zero points in the $r$ band. We measured the total (asymptotic) magnitudes of the galaxies, along with the optical radii $r_{25}$ at the isophote 25~mag/arcsec$^2$, using their curves of growth based on the performed {\tt{IRAF}}/{\tt{ELLIPSE}} fitting. 

As one can see, both galaxies are classified as SB0. The two galaxies do not differ significantly in luminosity and represent quite modest by mass and size disc galaxies \citep[see e.g.][]{2003MNRAS.343..978S} with possible recent star formation: \citet{2007ApJS..173..619K} showed that early-type galaxies with NUV$-r<5.5$ might have experienced recent star formation (both NGC\,4452 and NGC\,4469 show colours close to this boundary value). However, NUV$-r<5.5$ may also be related to the presence of evolved hot stars, the UV-upturn \citep{1990ApJ...364...35G}.

We also present the mean colours $(g-z)$ for the inner region, outlined by an isophote of 24\,mag/arcsec$^2$, and the outer region between the isophotes 24\,mag/arcsec$^2$ and 26\,mag/arcsec$^2$ (columns (9) and (10) in Table~\ref{tab:main.tab}). As one can see, the outer region is slightly bluer than the inner one for both galaxies, but this difference is insignificant. 

\begin{table*}
\caption{Main parameters of the selected galaxies.}
\label{tab:main.tab}
\centering
    \begin{tabular}{ccccccccccccc}
    \hline
    \hline\\[-1ex]    
 Galaxy  & RA & DEC & D & Type & $r_{25}$ & $r_{25}$ & $m_r$ & $M_r$ & NUV$-r$ & log\,$M_{*}$ & $\langle g-z \rangle_\mathrm{o}$ & $\langle g-z \rangle_\mathrm{i}$   \\[+0.5ex]
    & (J2000) & (J2000) & (Mpc) & & (arcmin) & (kpc) & (mag) & (mag) & (mag) & ($M_\odot$) & (mag) & (mag) \\[+0.5ex]
    &         &         &  (1)  & (2) & (3) & (4) & (5) & (6) & (7) & (8) & (9) & (10)     \\[+0.5ex]
    \hline\\[-0.5ex]
NGC\,4452 & 12:28:43 & 11:45:18 & 16.8 & SB0  & 2.07 & 10.10& 11.62 & -19.51 & 5.32 & 9.95 & $1.15\pm0.07$ & $1.24\pm0.16$ \\[+0.5ex]
NGC\,4469 & 12:29:28 & 08:44:59 & 16.8 & SB0-a& 2.94 & 14.30& 10.85 & -20.28 & 5.49 & 10.46& $1.29\pm0.11$ & $1.32\pm0.34$ \\[+0.5ex]

    \hline\\[-0.5ex]
    \end{tabular}

   \parbox[t]{160mm}{ Coumns: \\
   (1) Distance taken from the NASA/IPAC Extragalactic Database (NED) which is based on \citet{1988cng..book.....T} and calculated using the Tully-Fisher method \citep{1977A&A....54..661T}. The method of surface brightness fluctuations \citep{1988AJ.....96..807T}, applied by \citet{2007ApJ...655..144M} for 79 galaxies of the Virgo cluster, gives the mean distance $D=16.5$~Mpc for ``all galaxies (no W' cloud)'', which is close the the distances adopted in this paper. \\
   (2) Morphological type from HyperLeda,  \\
   (3), (4) semi-major axis of the isophote 25\,mag/arcsec$^2$ in the $r$ band, \\
   (5), (6) asymptotic magnitude in the $r$ band taking into account the Galactic extinction from \citet{2011ApJ...737..103S}, \\
   (7) colour based on the GALEX and SDSS photometry (corrected for the Galactic extinction but not corrected for the internal dust attenuation), \\
   (8) total stellar mass computed from the galaxy luminosity in the \textit{WISE} W1 band and using the prescriptions from \citet{2013MNRAS.433.2946W},\\
   (9) mean colour for the region between the isophotes 24\,mag/arcsec$^2$ and 26\,mag/arcsec$^2$ (corrected for the Galactic extinction but not corrected for the internal dust attenuation),\\
   (10) mean colour for the region within the isophote 24\,mag/arcsec$^2$ (corrected for the Galactic extinction but not corrected for the internal dust attenuation).
}
\end{table*}

\subsection{Description of each galaxy}
\label{methods}

Below we consider each of the selected galaxies in detail.

We compare the created deep images for NGC\,4452 (SDSS, Legacy) and NGC\,4469 (HERON, SDSS, Legacy) and conclude that both galaxies show a tilted outer structure with respect to the inner one. Below we show the stacked images, isophote and colour maps, as well as the results of the isophote analysis, based on the Legacy data. We use the stacked SDSS images to perform multicomponent decomposition, mainly to derive the parameters for the outer component, which is of great interest in the current study. The HERON image for NGC\,4469 can be found in fig.~A1 in \citet{ 2020MNRAS.494.1751M}. 

Both galaxies show the same difference between the total magnitude at the isophote 25~mag/arcsec$^2$ and 28~mag/arcsec$^2$, according to their curves of growth, as 0.09\,mag, or an increase of 8.6\% of the flux within the optical radius.

\subsubsection{NGC\,4452}
\label{NGC4452}

\begin{table}
\caption{Results of the {\tt{IMFIT}} fitting NGC\,4452 for the stacked SDSS image. In parentheses we list the {\tt{IMFIT}} functions which we use for describing each galaxy component. The fixed parameters are marked by *.}
\label{tab:NGC4452_decomp.tab}
\centering
    \begin{tabular}{cccc}
    \hline
    \hline\\[-1ex]    
    Component & Parameter &  Value & Units \\[0.5ex]
    \hline\\[-0.5ex]
    1. Nucleus:        & PA$^{*}$ &  $90.0$ & deg  \\[+0.5ex]
    ({\it S\'ersic}) & $\epsilon^{*}$  &  $0.25$ &  \\[+0.5ex]
                     & $n^{*}$  &  $0.68$  &  \\[+0.5ex]
                     & $\mu_\mathrm{e,r}$ & $16.53\pm0.56$ & mag/arcsec$^2$  \\[+0.5ex]
                     & $r_\mathrm{e}^{*}$ & $0.11$ & arcsec \\[+0.5ex]  
                     & $r_\mathrm{e}^{*}$ & $0.009$ & kpc \\[+0.5ex]
                     & $f$ & $0.001\pm0.0003$ &  \\[+0.5ex]  
    2. Bulge:        & PA$^{*}$ &  $90.0$ & deg  \\[+0.5ex]
    ({\it S\'ersic}) & $\epsilon^{*}$  &  $0.7$ &  \\[+0.5ex]
                     & $n^{*}$  &  $1.06$  &  \\[+0.5ex]
                     & $\mu_\mathrm{e,r}$ & $19.00\pm0.40$ & mag/arcsec$^2$  \\[+0.5ex] 
                     & $r_\mathrm{e}^{*}$ & $2.10$ & arcsec \\[+0.5ex]  
                     & $r_\mathrm{e}^{*}$ & $0.17$ & kpc \\[+0.5ex]
                     & $f$ & $0.019\pm0.008$ &  \\[+0.5ex]    
    3. Bar:        & PA$^{*}$ &  $90.0$ & deg  \\[+0.5ex]
    ({\it S\'ersic\_GenEllipse}) & $\epsilon^{*}$  &  $0.95$ &  \\[+0.5ex]
                     & $C_0$&  $0.42\pm0.33$  &  \\[+0.5ex]
                     & $n^{*}$  &  $0.18$  &  \\[+0.5ex]
                     & $\mu_\mathrm{e,r}$ & $19.12\pm0.04$ & mag/arcsec$^2$  \\[+0.5ex]
                     & $r_\mathrm{e}^{*}$ & $11.98$ & arcsec \\[+0.5ex]   
                     & $r_\mathrm{e}^{*}$ & $0.97$ & kpc \\[+0.5ex]
                     & $f$ & $0.055\pm0.001$ &  \\[+0.5ex]  
    4. Lens:      & PA$^{*}$ &  $90.0$ & deg  \\[+0.5ex]
({\it S\'ersic\_GenEllipse}) & $\epsilon^{*}$  &  $0.92$ &  \\[+0.5ex]
                     & $C_0$&  $4.93\pm2.35$  &  \\[+0.5ex]
                     & $n^{*}$  &  $0.20$  &  \\[+0.5ex]
                     & $\mu_\mathrm{e,r}$ & $19.75\pm0.04$ & mag/arcsec$^2$  \\[+0.5ex]
                     & $r_\mathrm{e}^{*}$ & $23.46$ & arcsec \\[+0.5ex]   
                     & $r_\mathrm{e}^{*}$ & $1.90$ & kpc \\[+0.5ex]
                     & $f$ & $0.223\pm0.004$ &  \\[+0.5ex]   
    5. Disc:       & PA &  $90.5\pm0.2$ & deg  \\[+0.5ex]
    ({\it S\'ersic\_GenEllipse}) & $\epsilon$  &  $0.76\pm0.03$ &  \\[+0.5ex]
                     & $C_0$&  $-0.56\pm0.35$  &  \\[+0.5ex]
                     & $n$  &  $0.80\pm0.27$  &  \\[+0.5ex]
                     & $\mu_\mathrm{e,r}$ & $20.62\pm0.30$ & mag/arcsec$^2$  \\[+0.5ex] 
                     & $r_\mathrm{e}$ & $26.42\pm1.48$ & arcsec \\[+0.5ex]
                     & $r_\mathrm{e}$ & $2.14\pm0.12$ & kpc \\[+0.5ex]
                     & $f$ & $0.398\pm0.159$ &  \\[+0.5ex] 
    6. Disc:       & PA &  $95.5\pm2.3$ & deg  \\[+0.5ex]
    ({\it S\'ersic\_GenEllipse}) & $\epsilon$  &  $0.66\pm0.01$ &  \\[+0.5ex]
                     & $C_0$&  $-0.72\pm0.11$  &  \\[+0.5ex]
                     & $n$  &  $0.94\pm0.33$  &  \\[+0.5ex]
                     & $\mu_\mathrm{e,r}$ & $23.29\pm0.50$ & mag/arcsec$^2$  \\[+0.5ex]
                     & $r_\mathrm{e}$ & $65.93\pm10.62$ & arcsec \\[+0.5ex] 
                     & $r_\mathrm{e}$ & $5.34\pm0.86$ & kpc \\[+0.5ex]
					  & $M_r$ & $-18.23\pm0.37$ & mag \\[+0.5ex]
                     & $f$ & $0.304\pm0.104$ &  \\[+0.5ex]
                     & $r_{28}$ & $3.83\pm0.19$ & arcmin  \\[+0.5ex]  
                     & $r_{28}$ & $18.62\pm0.93$ & kpc  \\[+0.5ex] 
    \hline\\[-0.5ex]
    \end{tabular}
\end{table}

\begin{figure*}
\label{fig:NGC4452_image}
\centering
\includegraphics[width=8cm]{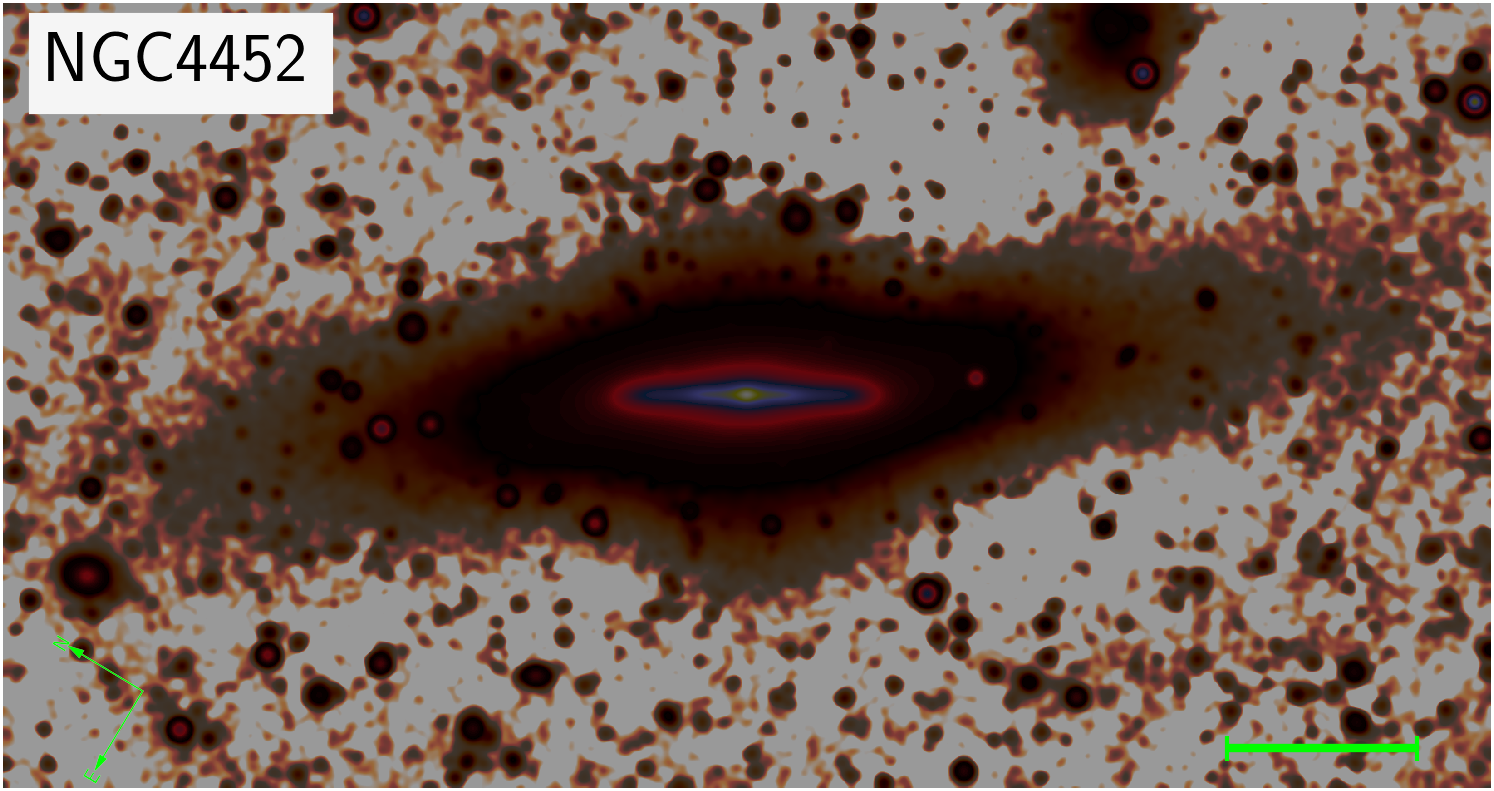}
\includegraphics[width=8cm]{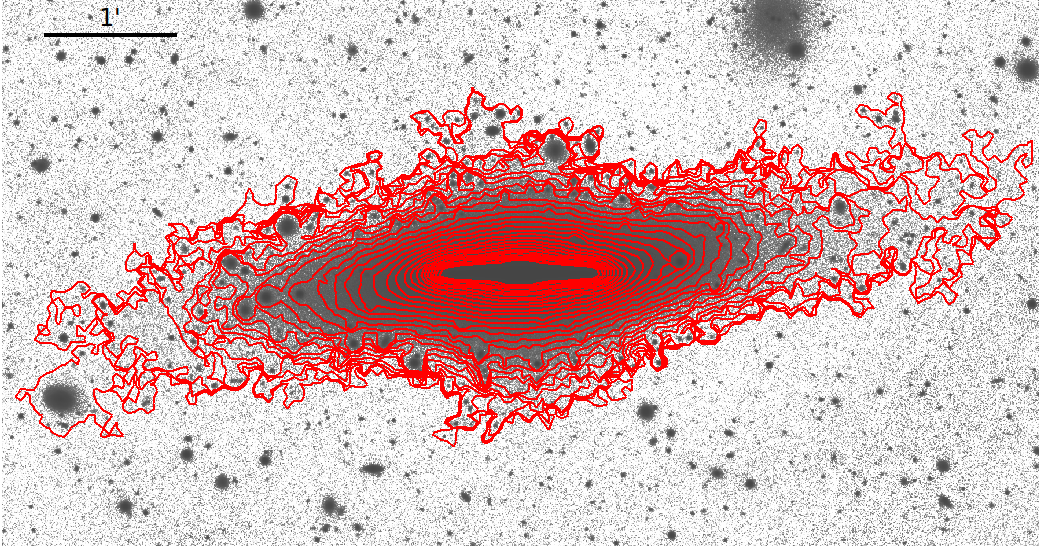}
\caption{Stacked Legacy $grz$ image for NGC\,4452 (left plot) and its superimposed isophotes from 20 to 26 mag/arcsec$^2$ (right plot)}
\end{figure*}

NGC\,4452 is an SB0 edge-on galaxy close to the centre of the Virgo cluster. \citet{2012ApJS..198....2K} classified it as S0c, according to their updated van den Bergh's \citep{1976ApJ...206..883V} parallel-sequence classification of galaxies.
In Fig.~\ref{fig:NGC4452_image}, one can see an inner disc of high surface brightness with a sharp edge. The very thin inner structure gives us a hint that this galaxy is exactly edge-on or very close to it. The outer disc exhibits a remarkable tilt and warping increasing towards the periphery. \citet{1994A&AS..105..481M} observe a twist of the isophotes which they attribute to a warped thick disc as in NGC\,4762. \citet{1994cag..book.....S} and \citet{2012ApJS..198....2K} also find NGC\,4452 similar to NGC\,4762. Our deep images do not reveal any interesting LSB features around NGC\,4452.

\begin{figure}
\label{fig:NGC4452_iraf}
\centering
\includegraphics[width=8.5cm]{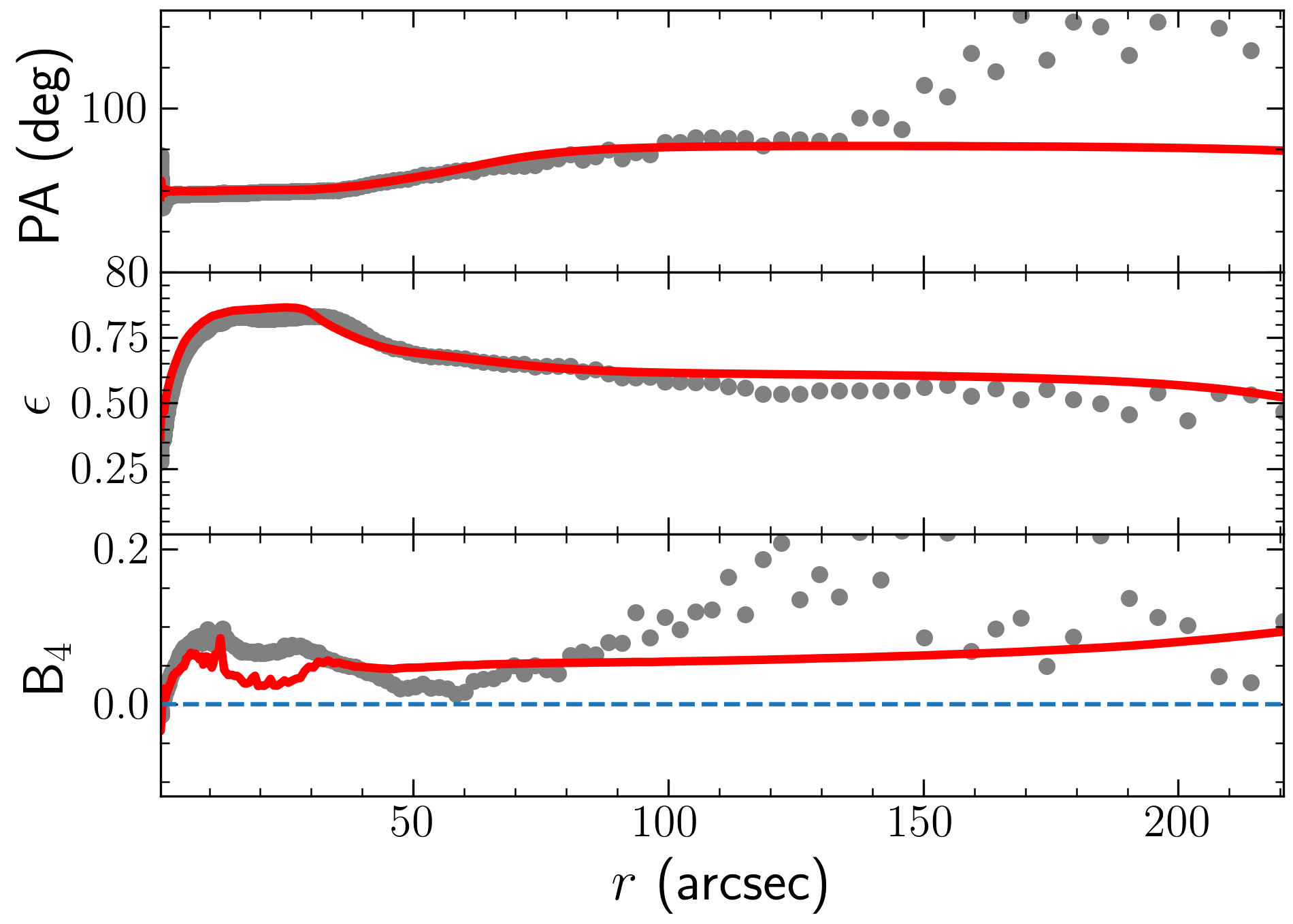}
\caption{Results of the {\tt{IRAF}}/{\tt{ELLIPSE}} fitting of NGC\,4452 for the stacked Legacy data. The galaxy image is rotated so that the plane of the inner structure is horizontal (the PA offset is $58.9\degr$). The red line shows the model.}
\end{figure}

In Fig.~\ref{fig:NGC4452_iraf}, one can see a steady increase of the position angle where the inner structure (disc or ring) is cut off: from $r\approx37\arcsec$ to the periphery the position angle changes up to $19\degr$. According to figure~8 in \citet{2012ApJS..198....2K}, who used a high-resolution HST ACS F475W image to study the photometry of this galaxy in great detail, the inner structure is very flat ($\epsilon\gtrsim0.9$), whereas due to the poor resolution of our stacked image the inner structure is smeared out and shows a lower flattening ($\epsilon\approx0.77$). The advantage of this work is that we can study the galaxy out to very low surface brightness where the outer structure becomes very thick (up to $\epsilon\approx0.5$ for $r\gtrsim190\arcsec$ in Fig.~\ref{fig:NGC4452_iraf}). The galaxy shows discy or oval isophotes at all radii ($B_4\gtrsim0$). Its outer structure is remarkably discy.

\begin{figure}
\label{fig:NGC4452_colour}
\centering
\includegraphics[width=8.5cm]{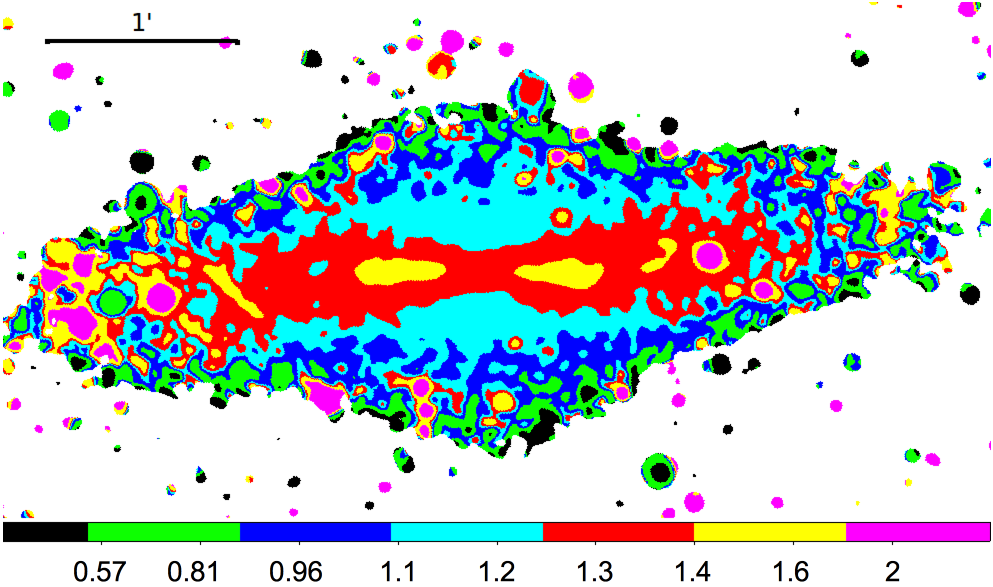}
\caption{Colour $(g-z)$ map based on the Legacy data for NGC\,4452 up to the isophote 26 mag/arcsec$^2$ in the $g$ band.} 
\end{figure}

\citet{2006ApJS..164..334F} find that there is no significant colour difference between the different components of the galaxy, except that its inner (nucleus) disc is bluer than the outer discs. Our lower-resolution colour map in Fig.~\ref{fig:NGC4452_colour} shows that the inner structure is red whereas the colour distribution above and below the galaxy plane is slightly bluer. Also, we note a prominent colour gradient (see Fig.~\ref{fig:1d_col_profiles} in Appendix~\ref{Appendix:1d_col_profiles}): the galaxy is getting bluer at larger radii from the centre. \citet{2016A&A...591A..38C} studied radial profiles of galaxies in the Coma and Virgo superclusters and found that early-type galaxies show no colour gradients. However, as no internal extinction correction was applied to NGC\,4452, its colour gradient may naturally arise from the non-negligible presence of dust which affects the colour distribution especially along the galaxy plane, though we see no sign of dust in this galaxy. Also, we can clearly see that the red inner disc has a prominent flaring and warping. The two more reddish regions, located symmetrically with respect to the galaxy centre in the plane (depicted by two \textit{yellow} ovals) probably point to a lens (a shelf-like feature in the surface brightness distribution) or, less likely, a ring structure.
As shown in many studies, inner rings are not common in SB0 galaxies, whereas bars, often
embedded in lenses of the same major-axis size, are often observed in such galaxies \citep{1961hag..book.....S,1979ApJ...227..714K,1982SAAS...12..115K,2007dvag.book.....B,2013seg..book....1K}. \citet{2014A&A...562A.121C} suggest, however, that about half of S0 galaxies 
have inner rings, though many of them do not have current or recent star formation \citep{2013A&A...555L...4C}. Also, lenses are seen in some galaxies that have no bars at all (e. g., NGC\,1553: \citealt{1975IAUS...69..367F,1984ApJ...286..116K,2013seg..book....1K}).
Overall, as there is no detection of H{\sc i} for this galaxy in HyperLeda (\citealt{2012MNRAS.422.1835S} give $\log\,$M(H{\sc i})$<7.27$, where masses are in Solar units), the star formation is probably very low and should be consistent with what we observe in regular S0 galaxies.

As emphasized in \citet{1979ApJ...227..714K} and based on the standard definitions and common
illustrations in \citet{1959HDP....53..275D,1961hag..book.....S,1982SAAS...12..115K,1994cag..book.....S,2007dvag.book.....B,2013seg..book....1K}, bars and lenses in early-type galaxies 1) both have very slowly decreasing surface brightnesses along their long axes (along
any axis, for a lens) and then a sharp outer edge (they represent ``shelves'' in their surface brightness profiles); 2) bars have higher surface brightnesses than the lenses in which they are commonly embedded; 3) most often, the bar exactly fills the lens along its longest dimension.
Using these facts, \citet{2012ApJS..198....2K} clearly distinguish five S\'ersic components in NGC\,4452: a nuclear stellar cluster ($n=0.7$), a small pseudobulge ($n=1.1$), a bar at a skew angle (not along the line of sight and not perpendicular to it) with $n=0.18$,  a lens component (with $n =0.2$), and an outer disc. In our decomposition we adopt the geometrical model by \citet{2012ApJS..198....2K} for the four inner (thin or tiny in our SDSS image) components (the nucleus, the pseudobulge, the bar, and the lens), implying that their geometry was well fitted by \citet{2012ApJS..198....2K} (due to the high resolution of the HST image used as compared to our SDSS image) and very sharp edges of the bar and the lens. Also, we assume that the geometry of these components does not change significantly for the HST ACS F475W and SDSS $r$ filters. In our fitting, we only fit the effective surface brightnesses and $C_0$ of these components. Note that we adopt the ellipticities of these components from \citet{2012ApJS..198....2K} (see their figure~8) and keep them fixed in our decomposition.  In contrast to the 1D fitting in \citet{2012ApJS..198....2K}, here we fit the overall 2D surface brightness distribution in NGC\,4452, -- this allows us to take into account the flattening, position angle and disciness/boxyness parameters for the remaining components -- an inner disc (which is especially visible in the colour map, see Fig.~\ref{fig:NGC4452_colour}, as a red colour component in the main galaxy plane) and a slightly bluer tilted outer disc. In the cases of the nucleus and the bulge, which are too tiny in our SDSS image to be well resolved, the $C_0$ parameter is set to 0, that is it implies pure elliptical isophotes. 

\begin{figure*}
\label{fig:NGC4452_decomp}
\centering
\includegraphics[width=18cm]{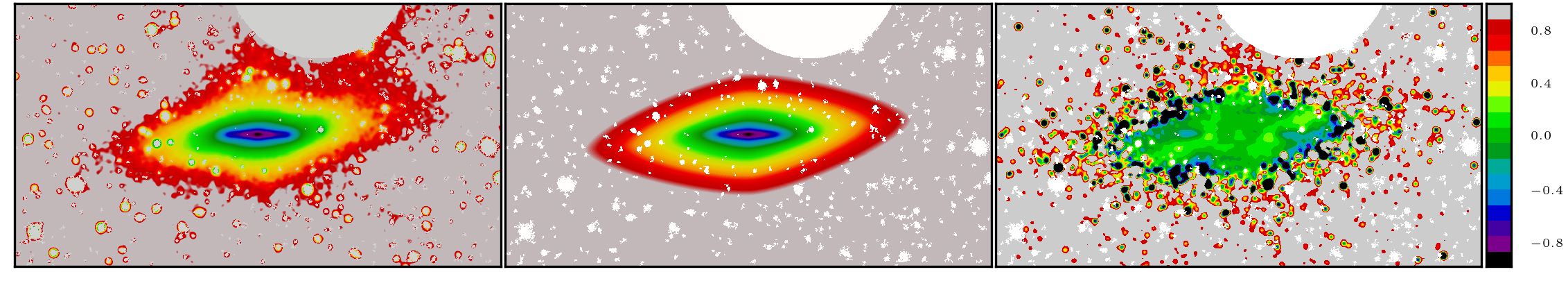}
\caption{Results of the photometric decomposition of NGC\,4452 for the stacked SDSS image: the observation (left), the model (middle) and the  residual image (right), which indicates the relative deviation between the fit and the image.}
\end{figure*}

In total, our model consists of six components, four of which have the fixed geometry as derived in \citet{2012ApJS..198....2K} and for the two outer components (discs) all parameters left free during the fitting. The results of our decomposition (see Table~\ref{tab:NGC4452_decomp.tab} and Fig.~\ref{fig:NGC4452_decomp}) confirm that the two outer components are indeed discs -- their S\'ersic indices are close to 1. The radial distribution in an edge-on transparent exponential disc is described by the function $r/h\cdot\mathrm{K}_1(r/h)$, where $h$ is the disc scalelength and $\mathrm{K}_1$ is the modified Bessel function of the second kind \citep{1981A&A....95..105V}. If one fits a S\'ersic function to this distribution, the S\'ersic index will be $n\approx0.8-0.9$ -- this is what we derive for both the discs in NGC\,4452. Also, their negative parameters $C_0$ show that these components have discy, diamond-like isophotes. Interestingly, the less extended disc coincides with the plane of the inner components, whereas the outer disc is inclined by $\sim8\degr$ with the plane of the inner structure. The outer tilted disc has a rather large contribution to the total galaxy luminosity ($\approx30$\%), despite the very low deprojected central surface brightness $\mu_{0,r}^\mathrm{face-on}=22.77$~mag/arcsec$^{-2}$ (for comparison, in the EGIS catalogue $\langle \mu_{0,r}^\mathrm{face-on} \rangle=21.56\pm0.81$~mag/arcsec$^{-2}$). However, its radial scalelength is also unusual ($h=3.51$~kpc) given its very low central surface brightness (see the correlation between the face-on central surface brightness and disc scalelength in figure~2 in \citealt{2010ApJ...722L.120F} and discussion in \citealt{2011ARA&A..49..301V}). It should also be noted that this disc has a very small flattening $\epsilon=0.66$ as compared to $\langle \epsilon \rangle=0.79\pm0.05$ for the EGIS galaxy discs. The inner disc, however, has typical characteristics for edge-on discs: $\langle \mu_{0,r}^\mathrm{face-on} \rangle=20.78$~mag/arcsec$^{-2}$, $h=1.75$~kpc, and $\epsilon=0.76$. 

In Fig.~\ref{fig:NGC4452_iraf} we superimpose the results of the {\tt{IRAF}}/{\tt{ELLIPSE}} fitting for our model. One can see that the model follows the observation fairly well. The increase of the position angle at large radii points to a warp of the outer tilted disc. Taking into account the above said, the outer disc can be considered as a tilted warped thick disc, whereas the inner disc is a warped and flared thin disc.

\subsubsection{NGC\,4469}
\label{NGC4469}

According to HyperLeda, this galaxy is classified as SB0-a. As NGC\,4452, it also belongs to the Virgo cluster, but it is located far out from its centre (given the distance to the cluster 16.5~Mpc, the projected distance from its centre to NGC\,4469 is 1.05~Mpc versus 0.24~Mpc for NGC\,4452). \citet{2016MNRAS.459.1276C} studied a prominent ``hump'' X-structure, which represents a manifestation of a side-on bar in edge-on galaxies \citep[see e.g.][]{1990A&A...233...82C,2000A&A...362..435L}. According to the results of their isophote fitting, it is obvious that they do not consider the outer structure of this galaxy (see their figure~A1): their isophotes do not extend deeper than $\approx24$~mag/arcsec$^2$. 

\citet{2015ApJS..216....9C} give a detail description of NGC\,4469 in their appendix and note slightly twisted outer isophotes with respect to its inner disc. However, we found that the name NGC\,4469 should be replaced there by NGC\,4569, as NGC\,4469 does not appear anywhere else in their study, whereas NGC\,4569 is truly considered.

\citet{2018ApJ...862...25J} detected an appreciable extraplanar H$\alpha$ and UV emission, which is associated with diffuse extraplanar ionized gas and extraplanar dust. This vertically extended dust can effectively scatter UV starlight and H$\alpha$ from H{\sc ii} regions located in the galactic plane. 

\citet{2016JAsGe...5..269H} measure a rather strong warp of the disc in NGC\,4469 by estimating the warp degree using the areas of an outer isophote and its fitted ellipse.

\begin{figure*}
\label{fig:NGC4469_image}
\centering
\includegraphics[width=8cm]{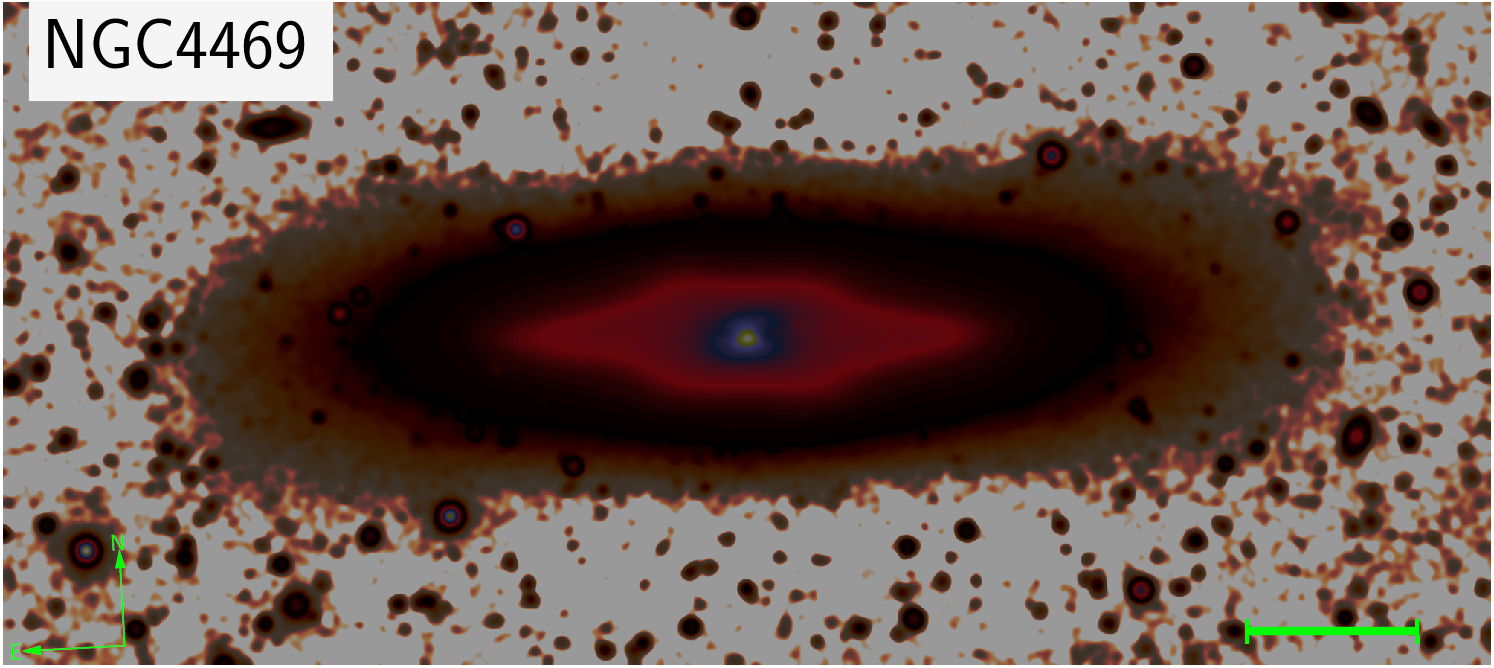}
\includegraphics[width=8cm]{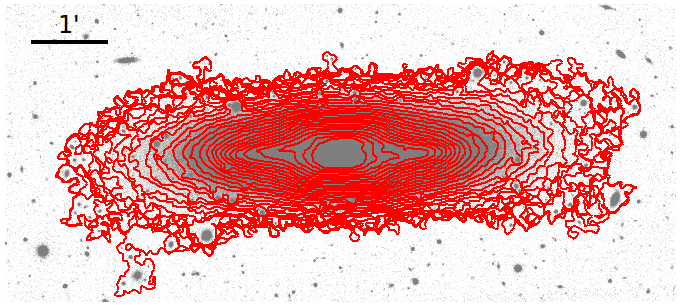}
\caption{Stacked Legacy $grz$ image for NGC\,4469 (left plot) and its superimposed isophotes from 20 to 26 mag/arcsec$^2$ (right plot).}
\end{figure*}

\citet{ 2020MNRAS.494.1751M} detect boxy/oval ($C_0=0.4$) isophotes for the outer structure in NGC\,4469. The HERON observation \citep{2019MNRAS.490.1539R} of this galaxy clearly shows that the outer structure is tilted by several degrees with respect to the inner one. Here, using the SDSS and Legacy imaging, we confirm their result. Fig.~\ref{fig:NGC4469_image} shows a thick boxy outer structure (a disc) which indeed has a prominent tilt. None of the deep observations, which we considered for this galaxy, does reveal any LSB details near NGC\,4469.
 
Its inclination $i$, based on a patchy dust structure in the galaxy centre, was estimated in \citet{ 2020MNRAS.494.1751M} as $88\degr$. However, here we re-estimated the inclination based on the GALEX NUV image, which, together with the optical images under study, shows a ring-like structure (it is also evident from a colour map, see below). For this ring we identified its abrupt edges and, based on its shape, we can use the formula (A1) $i=\mathrm{arccos}(\Delta z/R)$ from \citet{2015MNRAS.451.2376M}, where $R$ is the radius of the observed ring-like structure and $\Delta z$ is the maximum projected height of the inclined structure above (or below) the galaxy plane. We measured these parameters to be $R=100.8\arcsec$ and $\Delta z=14.4\arcsec$ and computed the galaxy inclination as approximately $82\degr$.

\begin{figure}
\label{fig:NGC4469_iraf}
\centering
\includegraphics[width=8.5cm]{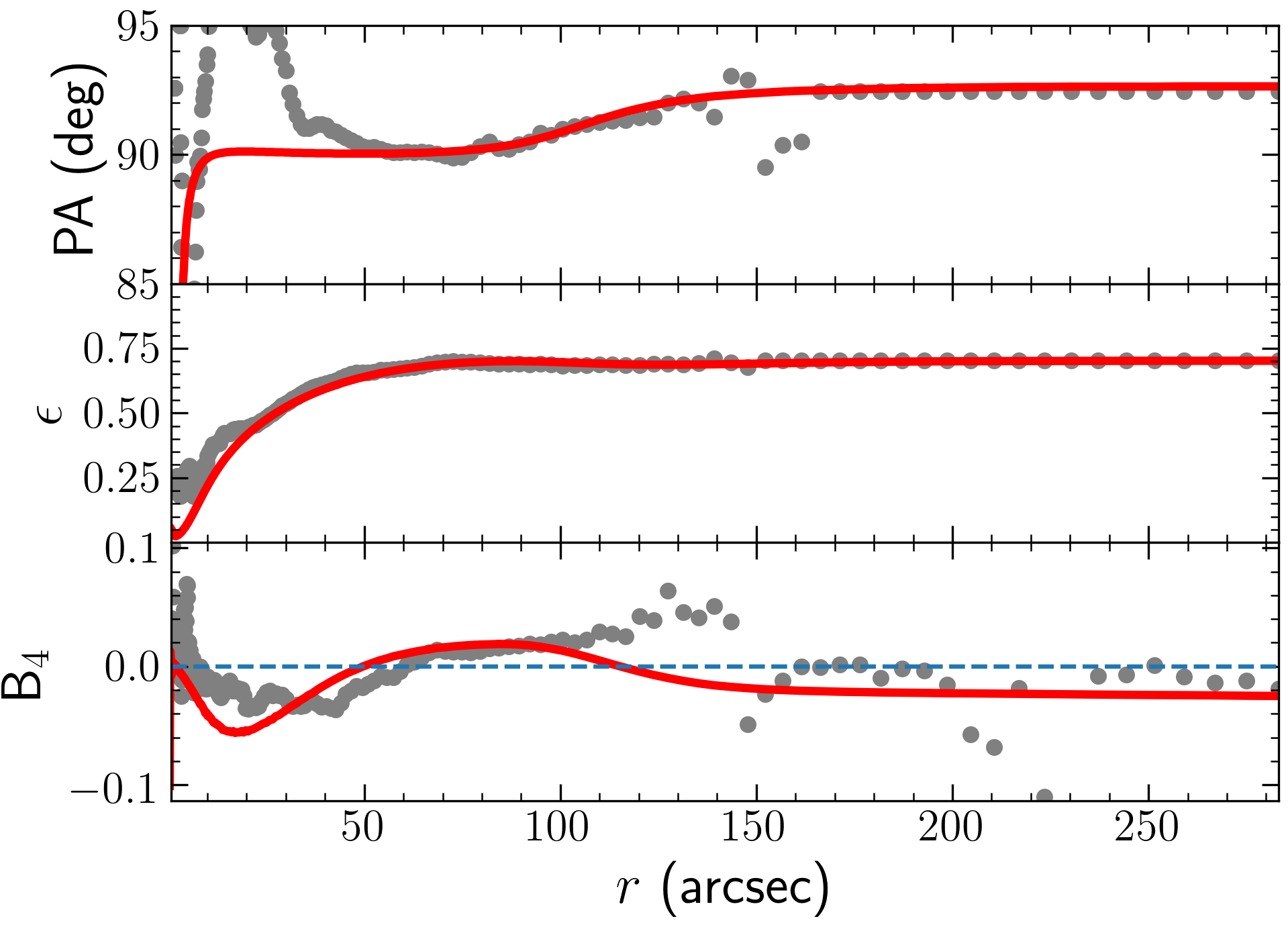}
\caption{Results of the {\tt{IRAF}}/{\tt{ELLIPSE}} fitting for NGC\,4469 for the stacked Legacy image. The galaxy image is rotated so that the plane of the inner structure is horizontal (the PA offset is $3.8\degr$). The red line shows the model.}
\end{figure}

The visual inspection of the results of the isophote fitting in Fig.~\ref{fig:NGC4469_iraf} reveals a change of the position angle of $\approx8\degr$, starting from the radius $75\arcsec$ up to the outermost isophotes. The inner region within $10\arcsec$ shows a round ($\epsilon\approx0$, $B_4\approx0$) component (bulge) which is surrounded by a prominent boxy (X-shape) bar ($B_4<0$ up to $r\approx70\arcsec$). At radii $75\arcsec$ to $150\arcsec$, we can see discy isophotes, which are probably related to a highly inclined inner ring. The slightly tilted outer structure at $r\gtrsim200\arcsec$ exhibits conspicuous boxy isophotes ($B_4<0$).

\begin{figure}
\label{fig:NGC4469_colour}
\centering
\includegraphics[width=8.5cm]{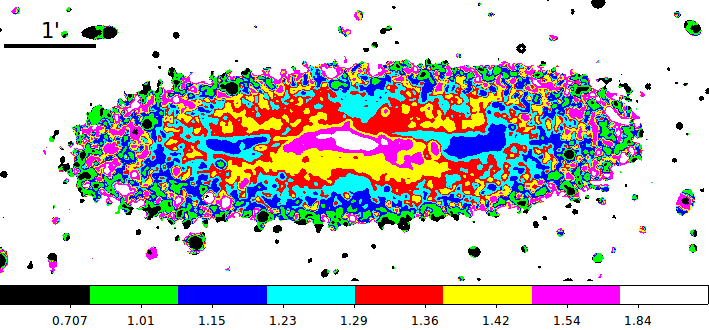}
\caption{Colour $(g-z)$ map based on the Legacy data for NGC\,4469 up to the isophote 26 mag/arcsec$^2$ in the $g$ band.}
\end{figure}

Our $(g-z)$ map in Fig.~\ref{fig:NGC4469_colour} obviously exhibits some traces of the patchy dust over the galaxy body (depicted by \textit{pink} and \textit{yellow} smudges), mainly in the central galaxy region. We can clearly see blue tips of an inner structure (possibly, a ring) and a reddish X-structure, surrounded by a blueish outer disc. The colour profile in Fig.~\ref{fig:1d_col_profiles} shows a very red central region (the internal extinction is one of the main reasons of this peak) and a steady blueing with radius where the disc dominates. The comparison with NGC\,4452 shows that the disc of NGC\,4469 is slightly redder. Nevertheless, as H{\sc i} mass is estimated to be $4.9 \times 10^7\mathrm{M}_{\sun}$ (calculated as $M_\mathrm{H{\sc I}}=2.36 \times 10^5 \times D^2 \times F_\mathrm{H{\sc I}}$, where the total flux $F_\mathrm{H{\sc I}}$ in Jy\,km\,s$^{-1}$ is taken from Table~1 in \citealt{2012MNRAS.423..787T}) and the emission lines in its SDSS spectrum are excited by star formation.   

\begin{figure*}
\label{fig:NGC4469_decomp}
\centering
\includegraphics[width=18cm]{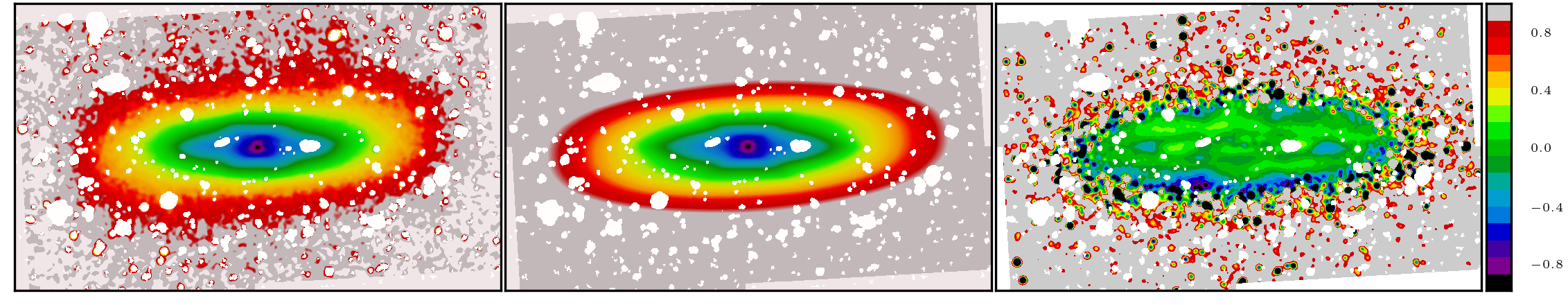}
\caption{Results of the photometric decomposition of NGC\,4469 for the stacked SDSS image: the observation (left), the model (middle) and the  residual image (right), which indicates the relative deviation between the fit and the image.}
\end{figure*}

Our decomposition model for this galaxy includes four primary components: bulge, bar, ring, and disc. For uniformity, each of them is described by a S\'ersic function. The results of the decomposition are listed in Table~\ref{tab:NGC4469_decomp.tab}. The goodness of the model is shown in Fig.~\ref{fig:NGC4469_decomp}. 
In Fig.~\ref{fig:major_axis_profiles}, we show a surface brightness profile with a superimposed decomposition model. 

According to our decomposition results, the bulge has an exponential profile ($n\approx1.3$) which makes it a pseudobulge (we admit, however, that due to internal dust extinction, the parameters of the bulge may be unreliable). The bright bar is significantly puffed up in the vertical direction ($\epsilon=0.3$). Its profile is close to exponential ($n\approx0.8$) and the isophotes are extremely boxy ($C_0=9$), which can be explained by the tremendous X-structure. Many or even most well developed bars have gone through a phase of vertical buckling instability that results in a boxy-shaped or even X-shaped inner structure \citep{1981A&A....96..164C,1990A&A...233...82C,1990ApJ...363..391P,1991A&A...252...75P,1991Natur.352..411R,2002MNRAS.330...35A,2005MNRAS.358.1477A,2006ApJ...637..214M,2019MNRAS.485.1900S}. The third component, which looks like a diamond-like disc, has a small S\'ersic index ($n\approx0.4$) and shows a smooth truncation at $\mu_\mathrm{r}=21.8$~mag/arcsec$^2$. We suppose that this component more closely resembles a ring (or, less likely, a tightly wound spiral). Its blue colour, unlike the reddish bar, shows that this component probably is not a lens \citep{2017A&A...599A..43H}. The outer disc, as in the case of NGC\,4452, is also close to exponential ($n\approx0.8$), but, in contrast, has boxy isophotes ($C_0=0.5$). Similar to NGC\,4452, its luminosity fraction $f$ is about 34\%. The disc has a very low surface brightness ($\mu_{0,r}^\mathrm{face-on}=23.11$~mag/arcsec$^{-2}$), while the radial extent is quite large for its luminosity (its scalelength $h=5.95$~kpc is larger than that of our Milky Way, see \citealt{2011ARA&A..49..301V} and references therein). According to our decomposition results, the difference between the position angles for the inner and outer structures is $3.5\degr$.

In Fig.~\ref{fig:NGC4469_iraf} we superimpose the results of the {\tt{IRAF}}/{\tt{ELLIPSE}} fitting for the model. As one can see, except for the central region, where the galaxy light is significantly affected by the dust attenuation (although the dusty patches were masked out during the fitting), the model follows the observation fairly well.

\begin{table}
\caption{Results of the {\tt{IMFIT}} fitting NGC\,4469 for the stacked SDSS image. In parentheses we list the {\tt{IMFIT}} functions which we use for describing each galaxy component. The fixed parameters are marked by *.}
\label{tab:NGC4469_decomp.tab}
\centering
    \begin{tabular}{cccc}
    \hline
    \hline\\[-1ex]    
    Component & Parameter &  Value & Units \\[0.5ex]
    \hline\\[-0.5ex]
    1. Bulge:        & PA$^{*}$ &  $90$ & deg  \\[+0.5ex]
    ({\it S\'ersic}) & $\epsilon^{*}$  &  $0.0$ &  \\[+0.5ex]
                     & $n$  &  $1.27\pm0.06$ &  \\[+0.5ex]
                     & $\mu_\mathrm{e,r}$ & $20.0\pm0.13$ & mag/arcsec$^2$  \\[+0.5ex]
                     & $r_\mathrm{e}$ & $4.69\pm0.49$ & arcsec \\[+0.5ex] 
                     & $r_\mathrm{e}$ & $0.38\pm0.04$ & kpc \\[+0.5ex]
                     & $f$ & $0.055\pm0.007$ &  \\[+0.5ex]    
    2. Bar:          & PA &  $89.9\pm1.1$ & deg  \\[+0.5ex]
    ({\it S\'ersic\_GenEllipse}) & $\epsilon$  &  $0.27\pm0.01$ &  \\[+0.5ex]
                     & $C_0$&  $9.2\pm3.1$  &  \\[+0.5ex]
                     & $n$  &  $0.77\pm0.22$  &  \\[+0.5ex]
                     & $\mu_\mathrm{e,r}$ & $21.17\pm0.11$ & mag/arcsec$^2$  \\[+0.5ex]
                     & $r_\mathrm{e}$ & $19.14\pm5.43$ & arcsec \\[+0.5ex] 
                     & $r_\mathrm{e}$ & $1.55\pm0.44$ & kpc \\[+0.5ex]
                     & $f$ & $0.221\pm0.053$ &  \\[+0.5ex]    
    3. Ring:      & PA &  $90.0\pm0.5$ & deg  \\[+0.5ex]
    ({\it S\'ersic\_GenEllipse}) & $\epsilon$  &  $0.79\pm0.01$ &  \\[+0.5ex]
                     & $C_0$&  $-0.4\pm0.1$  &  \\[+0.5ex]
                     & $n$  &  $0.38\pm0.07$  &  \\[+0.5ex]
                     & $\mu_\mathrm{e,r}$ & $21.27\pm0.05$ & mag/arcsec$^2$  \\[+0.5ex]
                     & $r_\mathrm{e}$ & $62.35\pm7.78$ & arcsec \\[+0.5ex] 
                     & $r_\mathrm{e}$ & $5.05\pm0.63$ & kpc \\[+0.5ex]
                     & $f$ & $0.387\pm0.086$ &  \\[+0.5ex]    
    4. Disc:      & PA &  $93.1\pm0.4$ & deg  \\[+0.5ex]
    ({\it S\'ersic\_GenEllipse}) & $\epsilon$  &  $0.69\pm0.02$ &  \\[+0.5ex]
                     & $C_0$&  $0.5\pm0.2$  &  \\[+0.5ex]
                     & $n$  &  $0.81\pm0.22$  &  \\[+0.5ex]
                     & $\mu_\mathrm{e,r}$ & $23.25\pm0.19$ & mag/arcsec$^2$  \\[+0.5ex]
                     & $r_\mathrm{e}$ & $90.25\pm2.47$ & arcsec \\[+0.5ex]                      
                     & $r_\mathrm{e}$ & $7.31\pm0.20$ & kpc \\[+0.5ex]
                     & $M_r$ & $-19.07\pm0.19$ & mag \\[+0.5ex]   
                     & $f$ & $0.337\pm0.035$ &  \\[+0.5ex]
                     & $r_{28}$ & $5.10\pm0.34$ & arcmin  \\[+0.5ex]
                     & $r_{28}$ & $24.78\pm1.67$ & kpc  \\[+0.5ex]      
    \hline\\[-0.5ex]
    \end{tabular}
\end{table}

\section{Discussion}
\label{sec:discussion}
In this section we discuss possible reasons of the observed phenomenon of the tilted discs in NGC\,4452 and NGC\,4469.

\subsection{Tilted disc or warps?}
\label{subsec:warps}

\begin{figure}
\label{fig:cenline}
\centering
\includegraphics[width=8cm]{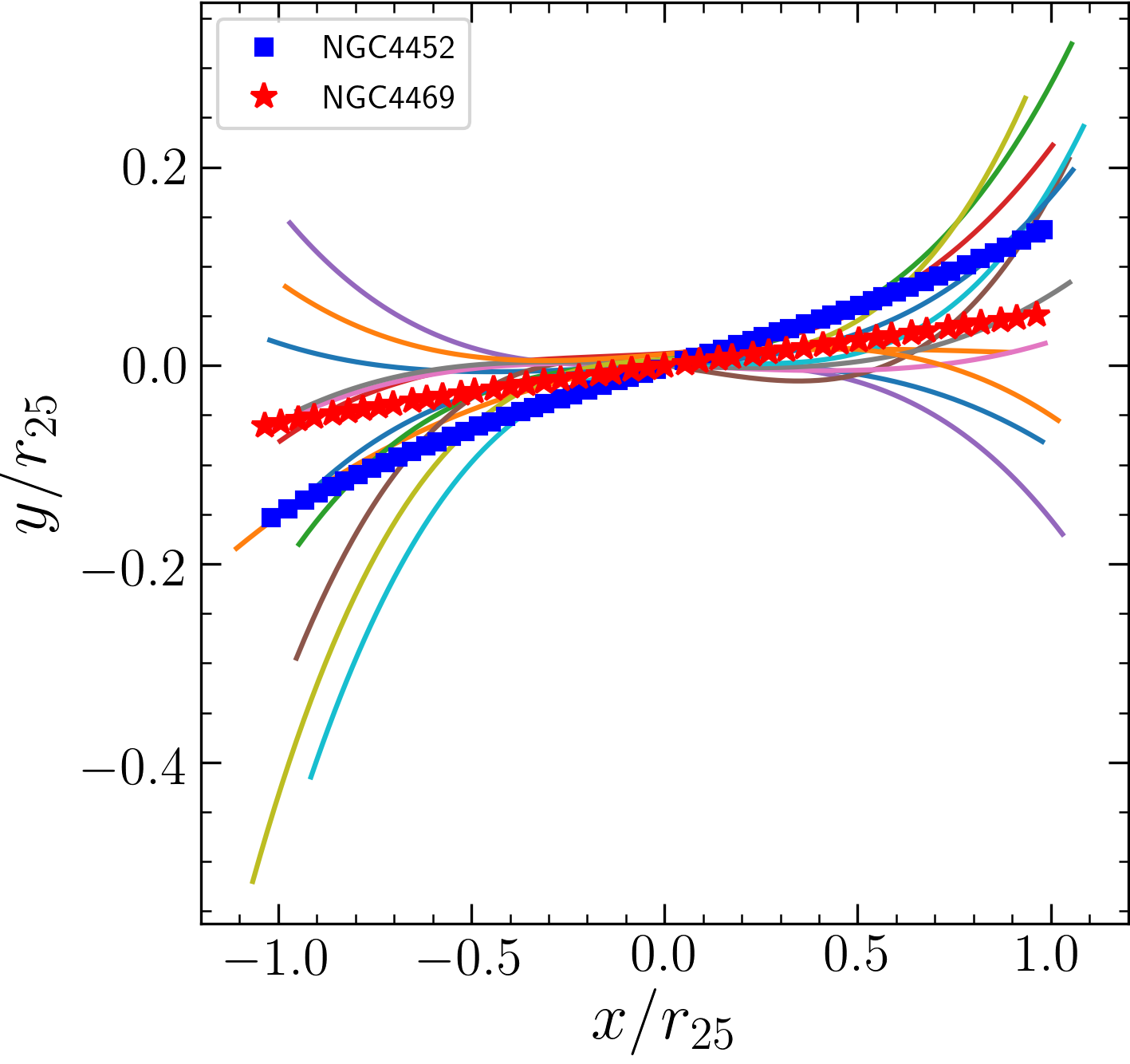}
\caption{Centre-lines for the galaxies with conspicuous warps from \citet{2016MNRAS.461.4233R} compared to the two galaxies with tilted outer structures, which we consider in the present paper. All centre-lines were plotted for isophotes of 25.5~mag/arcsec$^2$ in the $r$ band and approximated by splines. The $x$ and $y$ dimensions for each galaxy are normalized by its optical radius $r_{25}$.}
\end{figure}

To show that the tilted outer structures in NGC\,4452 and NGC\,4469 do not resemble ordinary warps of galactic discs, we collected the centrelines of 13 galaxies with conspicuous optical warps from \citet{2016MNRAS.461.4233R}. These centrelines were created for isophotes of 25.5~mag/arcsec$^2$ in the $r$ band. For the galaxies, which we study in this paper, we also created centrelines for the same isophote (see Fig.~\ref{fig:warps}). Prior to that, their stacked images were interpolated in the masked regions. As one can see in Fig.~\ref{fig:cenline}, the centre-lines of the outermost isophotes for the galaxies with tilted outer structures are virtually straight lines (their position angles are constant with radius), except that NGC\,4452 shows some warping far from the galaxy centre, whereas the centre-lines for the galaxies with warps are appreciably curved. In Fig.~\ref{fig:warps}, we pay your attention that the minor axes of the outermost isophotes for both galaxies are not perpendicular to their general galaxy planes (major axes of the inner isophotes).  

A more straightforward way to compare the warps from \citet{2016MNRAS.461.4233R} and the ``warps'' in NGC\,4452 and NGC\,4469 is to compare the parameters of their warps, the warp angle $\Psi$ (an angle measured between the galaxy plane and the line connecting the galaxy centre and the tips of the outer 25.5 isophote) and the radius $r_\mathrm{w}$ where the warp begins (see Fig.~\ref{fig:sketch}).  We created centre-lines for each of the isophotes plotted and then averaged them. That resulted in a final centre-line which was then fitted with a double piecewise linear function (see \citealt{2016MNRAS.461.4233R} for details). In Fig.~\ref{fig:warps} by red lines we show the centre-lines for each of the galaxies. For NGC\,4452, we obtained $r_\mathrm{w}=(0.18\pm0.02)\,r_{25}$, $\Psi_\mathrm{NW}=-7.6\degr$ (north-west, or left warp), $\Psi_\mathrm{SE}=10.2\degr$ (south-east, or right warp). For NGC\,4469, $r_\mathrm{w}=(0.60\pm0.03)\,r_{25}$, $\Psi_\mathrm{S}=-5.5\degr$ (south, or left warp), $\Psi_\mathrm{N}=5.6\degr$ (north, or right warp). For comparison, in \citet{2016MNRAS.461.4233R}, $\langle \Psi \rangle=7.3\pm6.4\degr$ and $\langle r_\mathrm{w} \rangle=(0.90\pm0.24)\,r_{25}$.  We can see that the warps in both galaxies start at smaller radii (especially in NGC\,4452) than observed in galaxies with prominent warps. Also, our models of NGC\,4452 and NGC\,4469, which do not imply warping of the outer discs but only their different position angles with respect to the inner components, describe the observations fairly well, including the major axis profiles, as well as PA, ellipticity and $B_4$ distributions.  
From this we can conclude that if the observed outer structures in NGC\,4452 and NGC\,4469 truly are genuine disc warps, they do not appear typical and deserve a special attention in any event.

\subsection{Triaxiality of the outer or inner structure?}
\label{subsec:triaxiality}
     
Another important issue regarding the tilted structures is the projection effect, when the galaxy inclination and a specific orientation of a triaxial component in this galaxy can give us an illusion that the inner or outer structure is tilted with respect to the outer or inner galaxy region, respectively. 

\begin{figure*}
\label{fig:inner_structure}
\centering
\includegraphics[width=8cm]{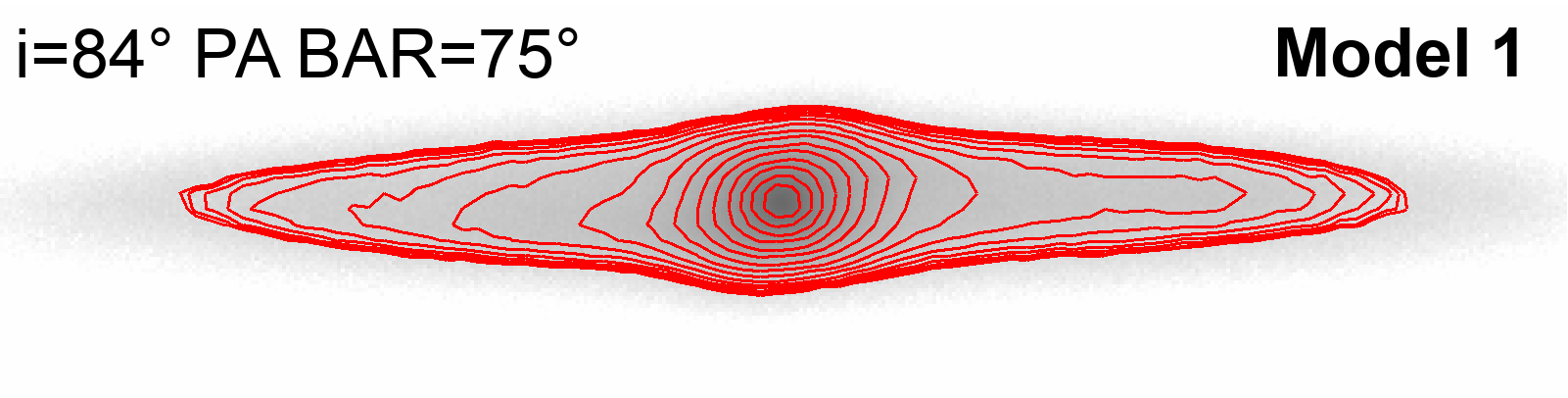}
\includegraphics[width=8cm]{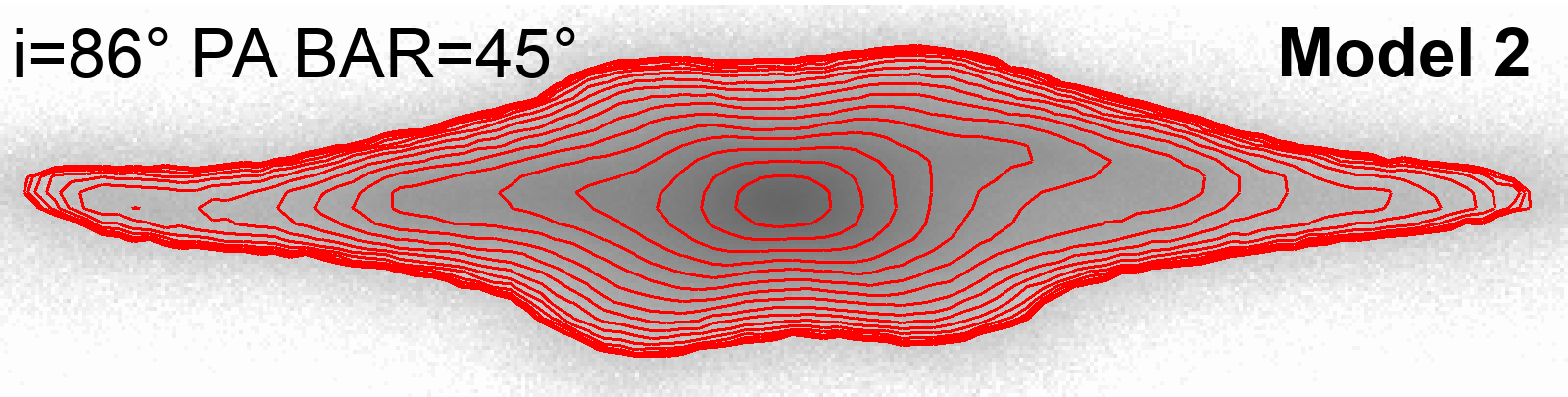}
\includegraphics[width=8cm]{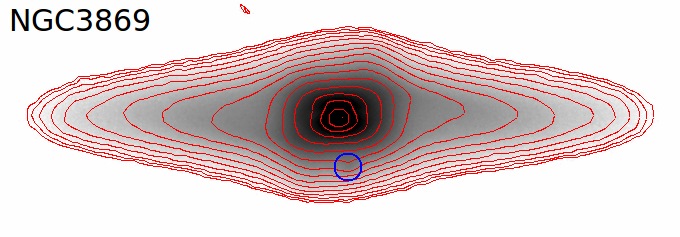}
\includegraphics[width=8cm]{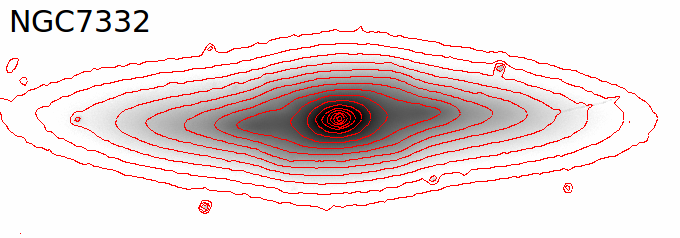}

\caption{\textit{Top panels:} Two different simulations of galaxies with the specific inclinations and position angles of the bar/bar plus the lens, which illustrate the effect of the ``inner tilting''. \textit{Bottom panels:} Real galaxies which demonstrate the same effect. The blue circle shows a satellite.}
\end{figure*}

As NGC\,4469 is not a pure edge-on galaxy and has a bright triaxial bar, the projection effect cannot be neglected. To show this, we performed numerical simulations to create mock galaxies with B/PS bulges (see details in the Appendix~\ref{Appendix:sim}). By changing the galaxy inclination angle and position angle of the bar (the angle between the line of nodes and the line of sight from the galaxy centre to the observer), we can indeed find a combination of these two angles, which gives us an appearance of a galaxy with a tilted inner structure (see Fig.~\ref{fig:inner_structure}, top lefthand panel). Interestingly, in the EGIS catalogue we found a galaxy, NGC\,3869 (see Fig.~\ref{fig:inner_structure}, bottom lefthand panel), which resembles the mock galaxy (Model~1) --- its inner structure looks tilted and, moreover, its X-shape structure seems asymmetric (note that in the image under the boxy B/PS bulge, there is also a satellite shown by the blue circle). The distributions of position angle, ellipticity and $B_4$ with radius for NGC\,3869 and the mock galaxy are presented in Fig.~\ref{fig:model_iso}. Both galaxies show very similar distributions of these quantities. Therefore, it is highly likely that the asymmetric view and the tilt of the inner structure (bar) in NGC\,3869 is, in fact, a consequence of a projection effect. A similar effect is observed in NGC\,7332, where we can see a B/PS bulge with relatively bright ansae (Fig.~\ref{fig:inner_structure}, bottom righthand panel). In another simulation of ours, Model~2, (Fig.~\ref{fig:inner_structure}, top righthand panel), a bar with a lens and bright ansea naturally arise in the course of the model evolution. By changing the galaxy inclination and position angle of the bar, we can again obtain a projection where the inner structure is inclined with respect to the disc. NGC\,7332 was noted by \citet{1993A&AS...98...29M} among other 10 galaxies with tilted inner structures. Most of these galaxies also harbour bars, therefore the observed tilting in these galaxies can probably be explained by the same effect of projection, as in the case of NGC\,3869 and NGC\,7332. For a pedagogical purpose, in Appendix\,\ref{Appendix:NGC509} we discuss another galaxy, NGC509 from the HERON sample. This galaxy was initially selected for this paper as a galaxy with a tilted outer structure, but after a careful investigation we rejected it because its orientation is far from edge-on.

As to NGC\,4469, apart from the B/PS bulge it has an inner ring-like structure. As this structure should, in principle, be almost round \citep[inner rings typically have axial ratios of $~0.8-0.85$, see e.g.][]{1995ApJS...96...39B,2013seg..book....1K,2014A&A...562A.121C}, the orientation of its major axis should not strongly depend on the galaxy inclination and the position angle of the bar.

We note, however, that not only bars, bulges and lenses can be triaxial structures \citep{1979ApJ...227..714K,2010A&A...521A..71M,2012A&AT...27..325S,2018A&A...609A.132C}, but also large-scale discs. For example, in the lenticular galaxy NGC\,5485 \citet{2016AstL...42..163S} found a non-circular stellar exponential disc with a highly noncircular stellar rotation. 
Also, she detected two wide elliptical stellar rings in the unbarred lenticular galaxy NGC\,502, which might be formed due to a dry minor merging. Galaxies with such oval distortions or elliptical discs may be not so rare.

The triaxiality of the stellar haloes is still under question. For example, haloes in hybrid semianalytic plus N-body models by \citet{2005ApJ...635..931B} are oblate \citep[see also][]{2008ApJ...680..295B}, whereas \citet{2010MNRAS.406..744C} find triaxial haloes in their N-body only simulations of Milky Way-mass galaxies (see also \citealt{2014ApJ...783...95B}). In recent large volume cosmological hydrodynamical simulation Illustris \citep{2014MNRAS.444.1518V}, \citet{2018MNRAS.479.4004E} found triaxial stellar haloes in galaxies with a wide range of the stellar halo fraction: $0.6\lesssim b/a \lesssim 1.0$ and $0.4 \lesssim c/a \lesssim 0.9$ (in their notation, $a$, $b$, $c$ are the major to minor principal axes of the inertia tensor for the stars). However, they concluded that the simulated halos can be fairly oblate, with median $\langle b/a\rangle\sim0.9$ and $\langle c/a \rangle \sim 0.5$. \citet{2019MNRAS.485.2589M} came to the same conclusion: they studied halo global properties in the Auriga cosmological magneto-hydrodynamical high-resolution simulations of Milky Way-mass galaxies \citep{2017MNRAS.467..179G}: most of the Auriga haloes appeared to be oblate spheroids as well. 

For the Milky Way, it has been well-established that the shape of the stellar halo is an oblate, not a triaxial spheroid \citep{2000ApJ...540..825Y,2003AJ....125.1958L,2008ApJ...673..864J,2011MNRAS.416.2903D}. Observationally, it is shown that the stellar haloes are moderately flattened spheroids ($c/a\sim0.6$ with a considerable range) with surface brightness distributions that are well described by a $r^{-\alpha}$ law, where the power slope $\alpha$ is usually found to be 2--4 \citep[see e.g.][]{2004MNRAS.352L...6Z,2017MNRAS.466.1491H}, supported by cosmological hydrodynamical simulations by \citet{2011MNRAS.416.2802F}. 

In conclusion, if we assume that the outer component in NGC\,4469 has a triaxial shape (although the observations show that the outer structures in disc galaxies are more likely to be oblate spheroids), whether it is a thick disc or a bright flattened exponential halo, its triaxiality may be responsible for the observed difference in the position angles of the inner and outer structures.

\subsection{Misalignment between angular momenta of the disc and dark matter halo}
\label{subsec:halo_galaxies_misalignment}
Now let us turn to some possible explanations of the observed phenomenon of the tilted outer structures if we suppose that they are real and not due to a projection effect.

There is every reason to expect \citep[see e.g.][]{1983IAUS..100..187D,1989MNRAS.237..785O,1991ApJ...377..365K,1992ARA&A..30...51B,1995ApJ...442..492D, 2006MNRAS.370....2S,2011ARA&A..49..301V} that the baryons that end up in the
inner parts of a galaxy disc get accreted with an angular momentum vector
(1) that is different from that of the dark matter that is accreted from the
      cosmic web and that lands farther out,
(2) that is different from that of baryonic gas that cools from the warm-hot
      intergalactic medium \citep[see also][]{2001ApJ...552..473D} and that lands in an outer disc, and
(3) that is different from that of any satellites made of dark matter and baryons
      that may be accreted at large radii into the outer disc and/or the dark halo.
Cosmological, hydrodynamic simulations show that angular momentum vectors of the gas and dark matter in
haloes tend to be misaligned \citep{2002ApJ...576...21V,2003ApJ...597...35C,2005ApJ...628...21S,2012ApJ...750..107S}.
All these processes can induce tilting of the outer structure with respect to the inner one.
Therefore, misalignments of main discs with outer discs whose inclinations reflect the gravity of misaligned outer dark matter haloes are certainly to be expected. 
However, as observationally we find few galaxies with tilted outer structures, these mechanisms should not be very important for inducing the tilts which can be directly observed (see also below).

\subsection{Dark matter halo tumbling}
\label{subsec:dark_matter}

As shown in a number of early \citep{1978MNRAS.183..779B,1987ApJ...319..575B,1988ApJ...327..507F,1992ApJ...399..405W,1995MNRAS.276..417J,1998MNRAS.296.1061T,2000ApJ...544L..87Y} and more recent works \citep{2013MNRAS.429.3316B,2016MNRAS.458.1559Z}, dark matter haloes are generally triaxial (see the review by \citealt{2017PhyU...60....3Z}). The flattened potential of triaxial haloes can induce oval distortions and non-circular velocities \citep{2007MNRAS.377...50H} which can be quite strong in dwarf \citep{2007ApJ...657..773V} and low-surface brightness galaxies \citep{2006MNRAS.373.1117H}. Its triaxiality can rapidly change over time \citet{2011MNRAS.416.1377V} --- this can impart a significant external torque on the galaxy disc. \citet{2009ApJ...703.2068D} performed N-body simulations and considered the inner part of the dark matter to be aligned with the disc, while the outer dark matter halo is misaligned and slowly tumbling. They found that the misaligned and tumbling outer halo causes a slowly changing external torque on the disc and that, in its turn, induces long-lived transient warps and tilting \citep[see also][]{1983IAUS..100..177T,1988MNRAS.234..873S,1989MNRAS.237..785O,1991ApJ...376..467K,1999ApJ...513L.107D,1999MNRAS.303L...7J}, bar instabilities \citep{2009ApJ...703.2068D} and spiral structure \citep[see also][]{2013MNRAS.431.1230K,2015MNRAS.451.2889K,2016MNRAS.461.2789H}. Therefore, this effect can be at play to explain the observed tilting in the galaxies under study. Unfortunately, this mechanism cannot be verified observationally. Also, the recent study by \citet{2019MNRAS.488.5728E} showed that the dark matter halo by itself does not play a significant role in the disc tilting.

\subsection{Accretion}
\label{subsec:accretion}

S0 galaxies are often thought to be red and dead systems without current star formation. However, this claim has been challenged and evaluated in multiple studies of S0 galaxies \citep[see e.g. recent studies by][]{2015BaltA..24..426K,2018A&A...620L...7S,2019AJ....158....5P,2019ApJS..244....6S,2020A&A...634A.102P}. In recent study, for example, \citet{2020A&A...634A.102P} considered the ring S0 galaxy NGC\,4513. They found that its ionized gas counterrotates the stellar component and, therefore, it was accreted. They concluded that its blue ring, demonstrating current star formation, is a result of tidal disruption of a massive gas-rich neighbour in the past, or it may be a consequence of a long star-formation event provoked by a gas accretion from a cosmological filament. Strongly inclined ionized gas discs, observed in many S0 galaxies \citep{2015AJ....150...24K,2019ApJS..244....6S}, is another evidence of the external gas supply. As \citet{2019ApJS..244....6S} claim, ``a crucial difference of the accretion regime in S0s with respect to spirals: the geometry of gas accretion in S0s is typically off-plane''.

Recent accretion of a gas-rich satellite in the case of NGC\,4452 seems to be impossible by two reasons. First, most galaxies in the central region of the Virgo cluster are gas-poor galaxies due to, for example, ram pressure stripping. Second, the velocity dispersion of galaxies in a cluster is very high (approximately, up to several thousands of km/sec). Therefore, a fast encounter does not result in merging, but in galaxy harassment (see Sect.~\ref{subsec:tidal}). 

In the case of NGC\,4469, which is located quite far from the Virgo centre, an accretion of one or a few gas-rich dwarfs might  happened several Gyr ago. One of such events can be responsible for a fairly blue ring in this galaxy. 

Another possible scenario to form tilted structures might be gas accretion via cosmological filaments when thin filaments come into a galaxy under some angle and make an inclined gaseous disc \citep{1996ApJ...461...55T,1998ApJ...506...93T,2005MNRAS.363....2K,2006MNRAS.368....2D} with ongoing star formation. However, in the presence of X-ray emitting, hot gas in the Virgo cluster (especially in its central part where NGC\,4452 resides), it is hard to explain how gas accretion via filaments can act in such harsh conditions.

\subsection{Tidally-induced tilts and warps}
\label{subsec:tidal}
We suppose that the most plausible explanation of the observed tilted discs in NGC\,4452 and NGC\,4469 is galaxy harassment when high-speed
encounters of galaxies in a cluster do not result in a merger but significantly affect the shape (and morphology) of the interacting galaxies which cause the discs to warp and tilt. Numerical and cosmological simulations support this mechanism \citep{2000ApJ...534..598V,2014ApJ...789...90K,2020arXiv200207022S}.

As \citet{2012ApJS..198....2K} noted, similar to NGC\,4762, the outer disc in NGC\,4452 is thick, warped, and tidally distorted. This could have been caused by a gravitational encounter with IC\,3381. Note that twists of outer isophotes, observed in some galaxies, can be the result of tidal effects (see e.g the case of NGC\,205 in the vicinity of M\,31, \citealt{1982SAAS...12..115K}). Therefore, if we had observed the same galaxies from the edge, their outer structure would have appeared tilted with respect to the inner one.

The tilt of the outer structure in NGC\,4469 can have the same reason. Galaxy harassment has plausibly heated the outer disc of NGC\,4469 (meanwhile the bar has almost inevitably heated the inner disc), with time now for all the heating to have phase-mixed around the galaxy, leaving the disc thicker and very likely tilted with respect to the inner disc. 

Circumstantial evidence of this mechanism in NGC\,4452 and NGC\,4469 can be the presence of disc antitruncations in both galaxies (see Fig.~\ref{fig:major_axis_profiles}). As shown in multiple studies \citep[see e.g.][]{2005ApJ...626L..81E, 2008AJ....135...20E}, disc antitruncations can be produced by galaxy interactions.

Why is the effect of the tilting so rare (only three galaxies, including NGC\,4638, have been noticed so far)? This is one of few studies \citep[see also][]{1993A&AS...98...29M,2020MNRAS.494.1751M} where we observationally confirm this phenomenon, whereas the disc flaring, warping and lopsidedness are often seen in edge-on galaxies and have been extensively studied for decades \citep{1976A&A....53..159S,1998A&A...337....9R,2009PhR...471...75J,2011ApJ...741...28C,2014A&A...567A.106L,2016MNRAS.461.4233R}. The main reason of the lack of observational evidence of tilted discs and haloes is that these outer structures are not well seen in regular images, whereas in deep observations we can better visualize faint structures, including tilted envelopes. We suppose that this phenomenon should be common in dense massive galaxy clusters. Apparently, the number of galaxies with tilted structures will increase, as more and more deep observations of galaxies become available.

\section{Conclusions}
\label{sec:conclusions}
In this paper, we have considered two moderate-luminosity SB0 galaxies with prominent tilted outer structures. Using different optical images (SDSS and Legacy) and by means of the stacking of the galaxy images in different bands, we were able to increase the depth of the resultant images up to 28.3~mag/arcsec$^2$ for the Legacy and 27.6 for the SDSS survey. We performed isophote fitting and a complex photometric decomposition for each galaxy. Based on the obtained results, we report that these galaxies plausibly show a real tilting of the outer structure with respect to the inner region (i.e. the outer and inner structures are oriented in different, tilted planes), which cannot be only explained by disc warps (though this effect can also be present). For NGC,4452 we obtained $\Delta \mathrm{PA}\approx6\degr$, whereas for NGC\,4469 the tilt is lower but still prominent ($\Delta \mathrm{PA}\approx3\degr$). The outer discs in these galaxies have completely different shapes (discy, or diamond-like, in NGC\,4452 and boxy in NGC\,4469).   

We propose that some combination of a single high speed encounter and the cumulative effect of galaxy harassment (in different proportions), which can distort and tilt the outer galaxy structure in a cluster, is a more plausible explanation of the observed phenomenon of disc tilting in both galaxies. Another important scenario can be misalignment of the triaxial (oblate) dark matter halo and its inner stellar disc simply because of how dark matter haloes grow out of the cosmic hierarchy. In addition to that, the tumbling of the dark matter halo might be enhanced by a fast encounter several Gyr ago, which would cause the outer disc to tilt. Also, for NGC\,4469, another explanation of the observed tilt, which cannot be rejected, unless a thorough kinematic study is carried out, is a possible triaxiality of an outer disc (or flat halo) in a highly inclined (but not purely edge-on) galaxy. In spite of the slightly blue colour of the tilted outer structures in both galaxies (NGC\,4469 also harbours a blueish inner structure, possibly, a ring), which can be evidence of ongoing star formation from the gas captured due to accretion of several gas-rich dwarfs several Gyr ago, this mechanism seems to be far less likely to form a tilted outer structure (especially for NGC\,4452) in the troublesome Virgo cluster where encounters are too fast for accretion to occur. 

In our future paper, we are about to study the nature of the main structural components in NGC\,4469 and NGC\,4452 using deep 21\,cm H{\sc i} observations, optical spectral observations, and kinematic measurements.

Using numerical simulations, we also showed that tilted inner structures may be well explained by a specific galaxy inclination, together with a specific orientation of a triaxial inner component (bar, bulge, lens) with respect to the observer.

\newpage
\section*{Acknowledgements}
We thank the anonymous reviewer for a thorough and constructive referee's report which helped to improve the paper.
This research has made use of the NASA/IPAC Infrared Science Archive (IRSA; \url{http://irsa.ipac.caltech.edu/frontpage/}), and the NASA/IPAC Extragalactic Database (NED; \url{https://ned.ipac.caltech.edu/}), both of which are operated by the Jet Propulsion Laboratory, California Institute of Technology, under contract with the National Aeronautics and Space Administration.  This research has made use of the HyperLEDA database (\url{http://leda.univ-lyon1.fr/}; \citealp{2014A&A...570A..13M}). 
This work is based in part on observations made with the {\it Spitzer} Space Telescope, which is operated by the Jet Propulsion Laboratory, California Institute of Technology under a contract with NASA. 
AVM and AAS express gratitude for the grant of the Russian
Foundation for Basic Researches number 19-02-00249.
RMR acknowledges financial support from his late father Jay Baum Rich.
VPR acknowledges financial support from the Russian Science
Foundation (grant no. 19-12-00145). 

Funding for the Sloan Digital Sky Survey IV has been provided by the Alfred P. Sloan Foundation, the U.S. Department of Energy Office of Science, and the Participating Institutions. SDSS-IV acknowledges
support and resources from the Center for High-Performance Computing at
the University of Utah. The SDSS web site is www.sdss.org.

SDSS-IV is managed by the Astrophysical Research Consortium for the 
Participating Institutions of the SDSS Collaboration including the 
Brazilian Participation Group, the Carnegie Institution for Science, 
Carnegie Mellon University, the Chilean Participation Group, the French Participation Group, Harvard-Smithsonian Center for Astrophysics, 
Instituto de Astrof\'isica de Canarias, The Johns Hopkins University, Kavli Institute for the Physics and Mathematics of the Universe (IPMU) / 
University of Tokyo, the Korean Participation Group, Lawrence Berkeley National Laboratory, 
Leibniz Institut f\"ur Astrophysik Potsdam (AIP),  
Max-Planck-Institut f\"ur Astronomie (MPIA Heidelberg), 
Max-Planck-Institut f\"ur Astrophysik (MPA Garching), 
Max-Planck-Institut f\"ur Extraterrestrische Physik (MPE), 
National Astronomical Observatories of China, New Mexico State University, 
New York University, University of Notre Dame, 
Observat\'ario Nacional / MCTI, The Ohio State University, 
Pennsylvania State University, Shanghai Astronomical Observatory, 
United Kingdom Participation Group,
Universidad Nacional Aut\'onoma de M\'exico, University of Arizona, 
University of Colorado Boulder, University of Oxford, University of Portsmouth, 
University of Utah, University of Virginia, University of Washington, University of Wisconsin, 
Vanderbilt University, and Yale University.

The Legacy Surveys consist of three individual and complementary projects: the Dark Energy Camera Legacy Survey (DECaLS; NOAO Proposal ID \# 2014B-0404; PIs: David Schlegel and Arjun Dey), the Beijing-Arizona Sky Survey (BASS; NOAO Proposal ID \# 2015A-0801; PIs: Zhou Xu and Xiaohui Fan), and the Mayall z-band Legacy Survey (MzLS; NOAO Proposal ID \# 2016A-0453; PI: Arjun Dey). DECaLS, BASS and MzLS together include data obtained, respectively, at the Blanco telescope, Cerro Tololo Inter-American Observatory, National Optical Astronomy Observatory (NOAO); the Bok telescope, Steward Observatory, University of Arizona; and the Mayall telescope, Kitt Peak National Observatory, NOAO. The Legacy Surveys project is honored to be permitted to conduct astronomical research on Iolkam Du'ag (Kitt Peak), a mountain with particular significance to the Tohono O'odham Nation.

NOAO is operated by the Association of Universities for Research in Astronomy (AURA) under a cooperative agreement with the National Science Foundation.

This project used data obtained with the Dark Energy Camera (DECam), which was constructed by the Dark Energy Survey (DES) collaboration. Funding for the DES Projects has been provided by the U.S. Department of Energy, the U.S. National Science Foundation, the Ministry of Science and Education of Spain, the Science and Technology Facilities Council of the United Kingdom, the Higher Education Funding Council for England, the National Center for Supercomputing Applications at the University of Illinois at Urbana-Champaign, the Kavli Institute of Cosmological Physics at the University of Chicago, Center for Cosmology and Astro-Particle Physics at the Ohio State University, the Mitchell Institute for Fundamental Physics and Astronomy at Texas A\&M University, Financiadora de Estudos e Projetos, Fundacao Carlos Chagas Filho de Amparo, Financiadora de Estudos e Projetos, Fundacao Carlos Chagas Filho de Amparo a Pesquisa do Estado do Rio de Janeiro, Conselho Nacional de Desenvolvimento Cientifico e Tecnologico and the Ministerio da Ciencia, Tecnologia e Inovacao, the Deutsche Forschungsgemeinschaft and the Collaborating Institutions in the Dark Energy Survey. The Collaborating Institutions are Argonne National Laboratory, the University of California at Santa Cruz, the University of Cambridge, Centro de Investigaciones Energeticas, Medioambientales y Tecnologicas-Madrid, the University of Chicago, University College London, the DES-Brazil Consortium, the University of Edinburgh, the Eidgenossische Technische Hochschule (ETH) Zurich, Fermi National Accelerator Laboratory, the University of Illinois at Urbana-Champaign, the Institut de Ciencies de l'Espai (IEEC/CSIC), the Institut de Fisica d'Altes Energies, Lawrence Berkeley National Laboratory, the Ludwig-Maximilians Universitat Munchen and the associated Excellence Cluster Universe, the University of Michigan, the National Optical Astronomy Observatory, the University of Nottingham, the Ohio State University, the University of Pennsylvania, the University of Portsmouth, SLAC National Accelerator Laboratory, Stanford University, the University of Sussex, and Texas A\&M University.

BASS is a key project of the Telescope Access Program (TAP), which has been funded by the National Astronomical Observatories of China, the Chinese Academy of Sciences (the Strategic Priority Research Program "The Emergence of Cosmological Structures" Grant \# XDB09000000), and the Special Fund for Astronomy from the Ministry of Finance. The BASS is also supported by the External Cooperation Program of Chinese Academy of Sciences (Grant \# 114A11KYSB20160057), and Chinese National Natural Science Foundation (Grant \# 11433005).

The Legacy Survey team makes use of data products from the Near-Earth Object Wide-field Infrared Survey Explorer (NEOWISE), which is a project of the Jet Propulsion Laboratory/California Institute of Technology. NEOWISE is funded by the National Aeronautics and Space Administration.

The Legacy Surveys imaging of the DESI footprint is supported by the Director, Office of Science, Office of High Energy Physics of the U.S. Department of Energy under Contract No. DE-AC02-05CH1123, by the National Energy Research Scientific Computing Center, a DOE Office of Science User Facility under the same contract; and by the U.S. National Science Foundation, Division of Astronomical Sciences under Contract No. AST-0950945 to NOAO.

This publication makes use of data products from the Wide-field Infrared Survey Explorer, which is a joint project of the University of California, Los Angeles, and the Jet Propulsion Laboratory/California Institute of Technology, funded by the National Aeronautics and Space Administration.

This study makes use of observations made with the NASA Galaxy Evolution Explorer. GALEX is operated for NASA by the California Institute of Technology under NASA contract NAS5-98034.


\section*{Data availability}
The data underlying this article will be shared on reasonable request to the corresponding author.



\bibliographystyle{mnras}
\bibliography{art} 




\appendix

\section{NGC\,509}
\label{Appendix:NGC509}
\begin{figure*}
\label{fig:NGC509_image}
\centering
\includegraphics[width=8cm]{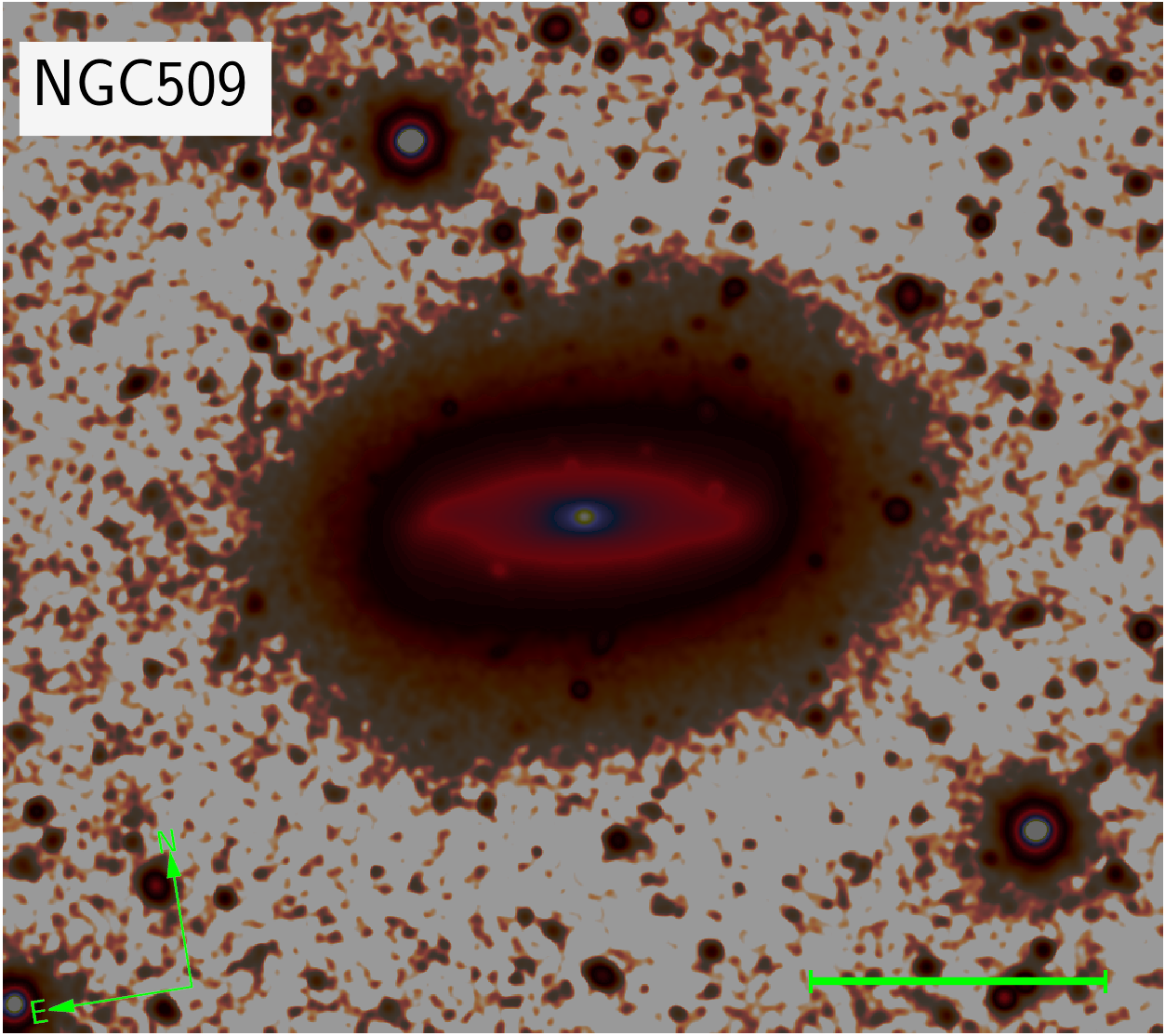}
\includegraphics[width=8cm]{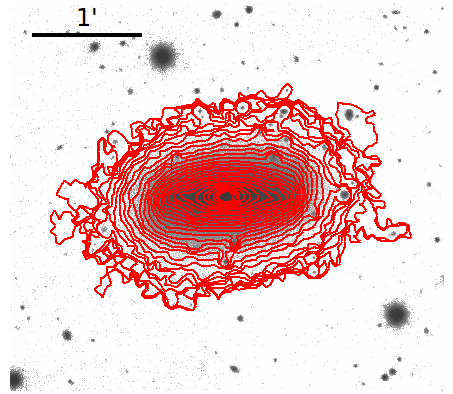}
\caption{Stacked Legacy $grz$ image for NGC\,509 (left plot) and its superimposed isophotes from 20 to 26 mag/arcsec$^2$ (right plot)}
\end{figure*}

Here we consider an initially selected ``edge-on'' galaxy with a tilted outer structure, NGC\,509. In Fig.~\ref{fig:NGC509_image} you can see its deep image which reveals an incredibly tilted halo (or thick disc) with respect to the stellar disc.

\begin{figure}
\label{fig:NGC509_contrast_enhanced}
\centering
\includegraphics[width=8cm]{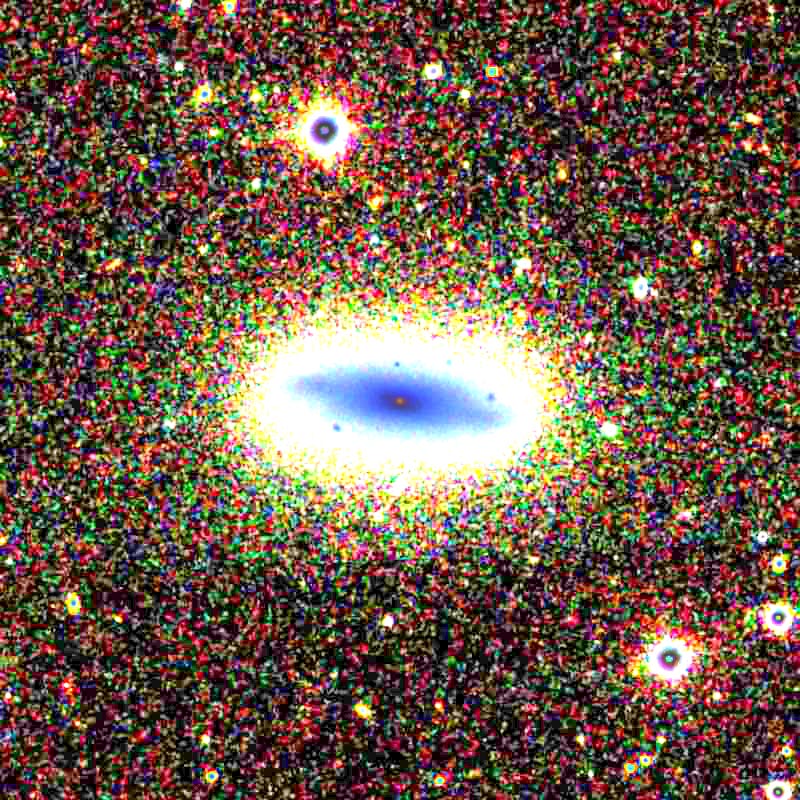}
\caption{The contrast-enhanced snapshot image of NGC\,509 based on a SDSS color image.}
\end{figure}

\begin{figure}
\label{fig:tilted-ring}
\centering
\includegraphics[width=8cm]{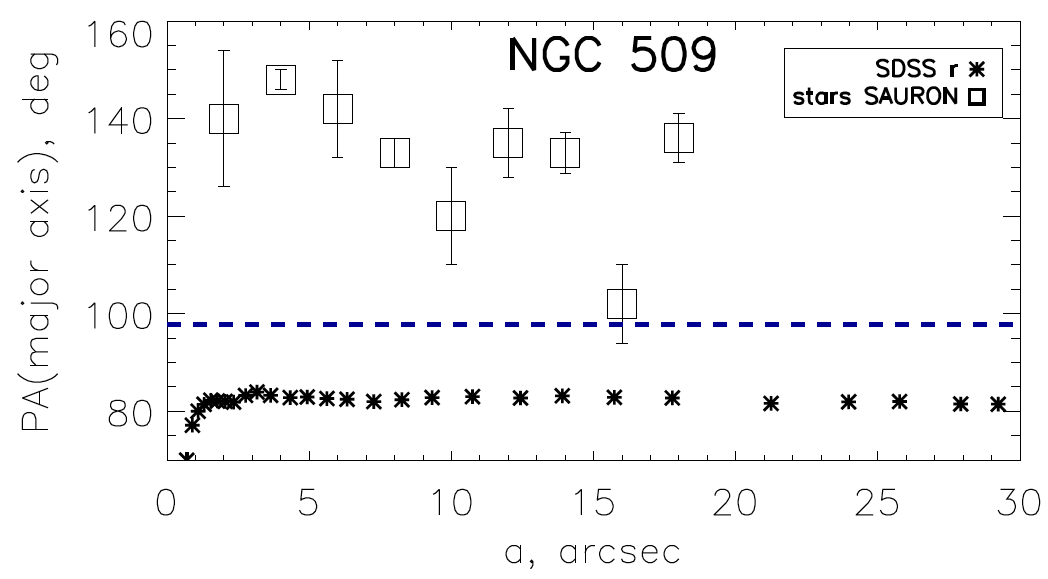}
\caption{The results of the tilted-ring analysis for the SAURON
stellar velocity field. The orientation of the kinematical major axis is
shown by large squares in comparison with the photometric major axis
(asterisks, the SDSS $r$ band) and with the outer disc line of nodes (blue
dashed line).}
\end{figure}

However, after a careful investigation, we concluded that NGC\,509 cannot be treated as an edge-on galaxy
with an inclined bright halo (or thick disc?). The kinematical tilted-ring analysis of
the 2D stellar velocity field, provided by the ATLAS-3D survey \citep{2011MNRAS.413..813C},
reveals that the disc inclination is equal to approximately $65\degr$.
It also reveals the bar-induced non-circular stellar rotation. The blue feature inside
the thin inner disc is probably a bar with blue ansae (see Fig.~\ref{fig:NGC509_contrast_enhanced}). The true
disc is seen beyond the bar, it has a photometrically implied orientation of the line of nodes equal to PA=$97\degr$. If we inspect
Fig.~\ref{fig:tilted-ring}, where we compare the photometric and kinematical major axes in the centre of NGC\,509 with the line of nodes for the outer disc (general plane of the galaxy), we can see that the photometric and kinematical major axes deviate from the line of nodes in different directions. Such a behaviour is predicted by dynamical simulations of star rotation in a non-axisymmetrical potential: the zero-velocity line, while in an axisymmetrical potential, traces the isophote minor axis, but in the bar-dominated potential it turns along the bar \citep{1997MNRAS.286..812V}. This means that the zero-velocity line tends to align with the major axis of the bar (photometric major axis of NGC\,509), whereas the kinematical major axis looks oriented perpendicular to the bar major axis. This proves that NGC\,509 is indeed a barred galaxy with a rather moderate inclination to the line of sight. Therefore its ``tilted'' outer structure is in fact a consequence of the galaxy inclination and the bar orientation with respect to the observer.

\section{Radial colour profiles}
\label{Appendix:1d_col_profiles}
\begin{figure*}
\label{fig:1d_col_profiles}
\centering
\includegraphics[height=7cm]{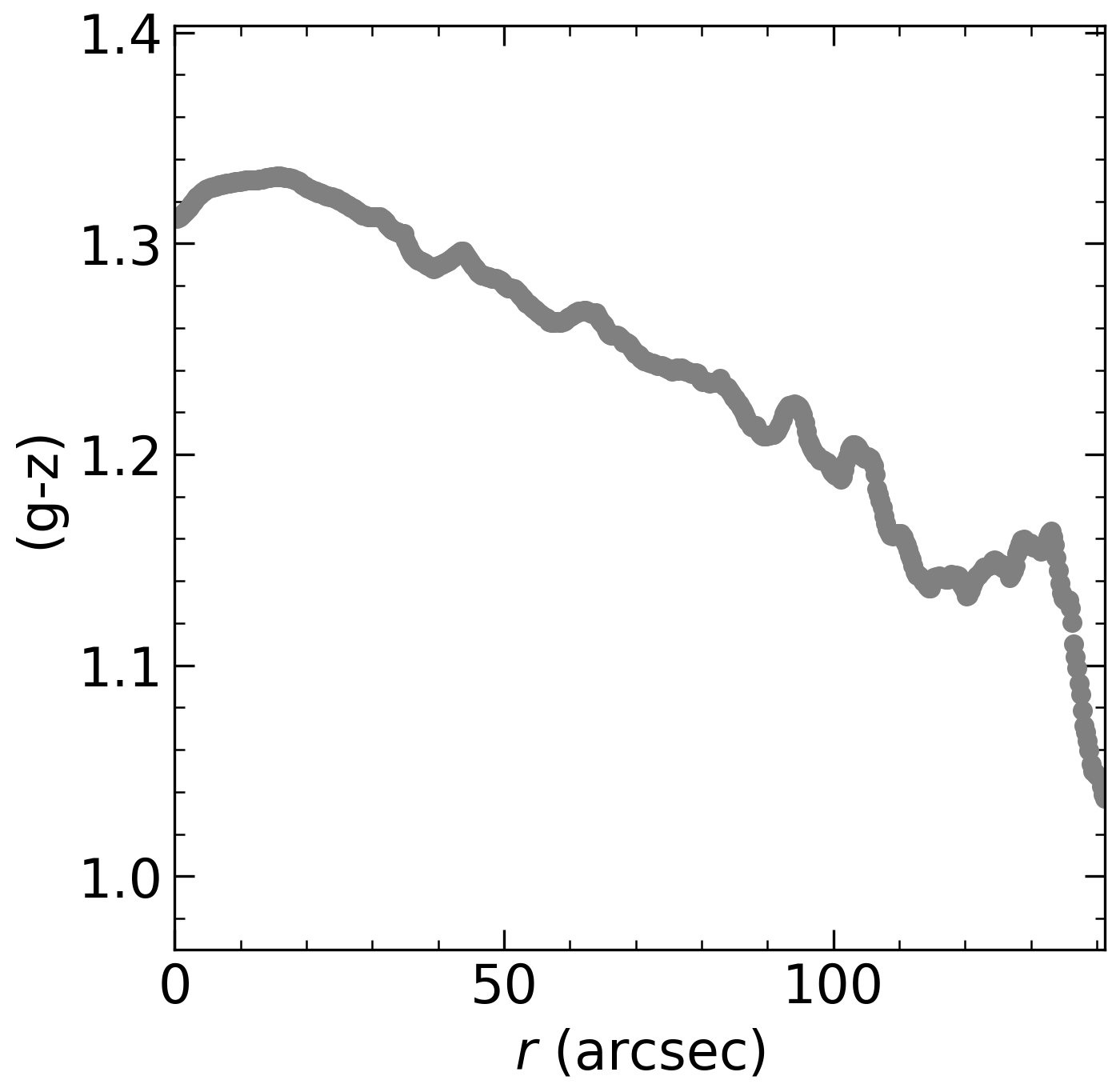}
\includegraphics[height=6.95cm]{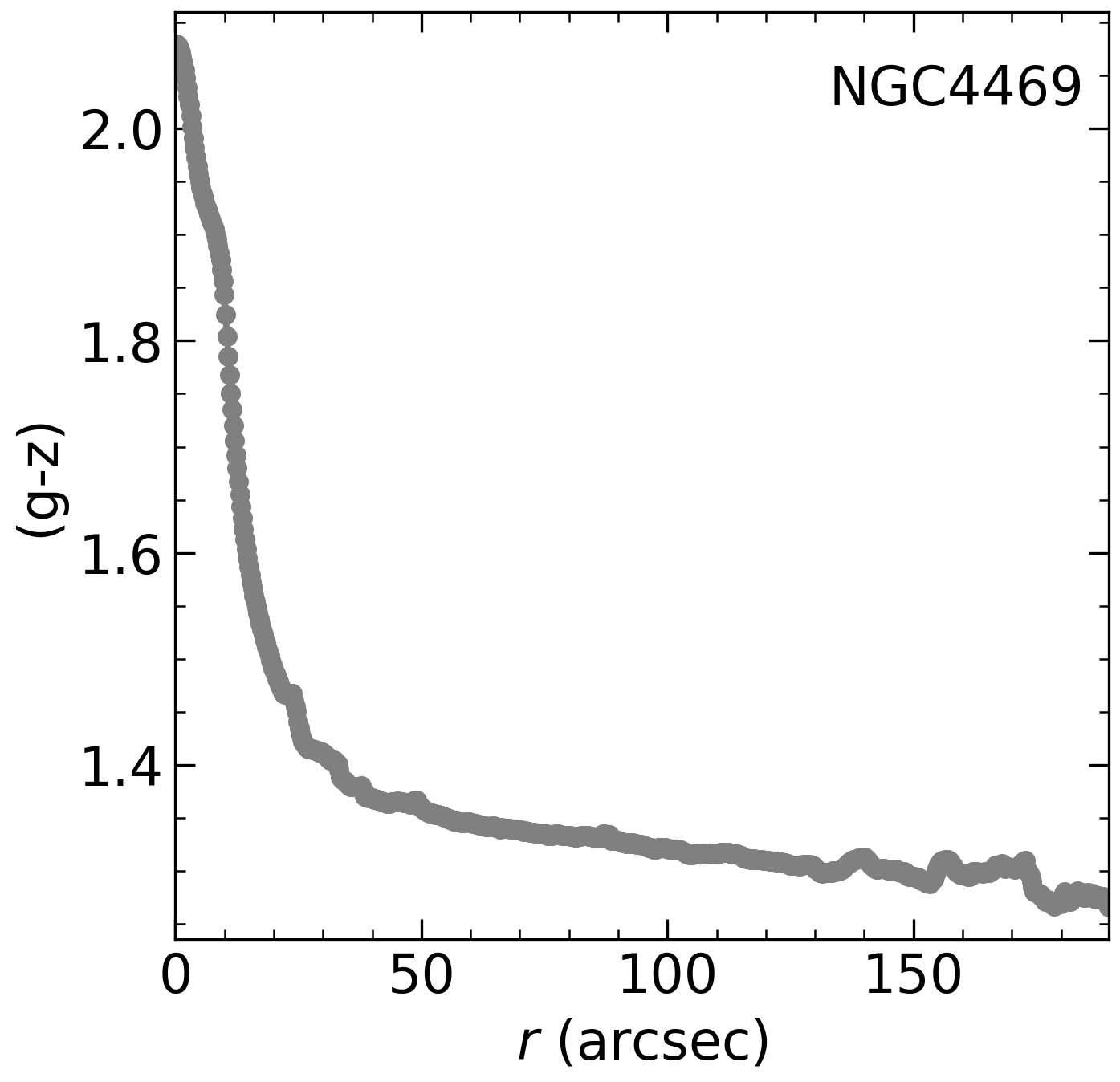}
\caption{Radial colour profiles for NGC\,4452 (the lafthand plot) and NGC\,4469 (right plot), which were created based on the Legacy data.}
\end{figure*}

\section{Major-axis profiles}
\label{Appendix:major_axis_profiles}
\begin{figure*}
\label{fig:major_axis_profiles}
\centering
\includegraphics[height=7cm]{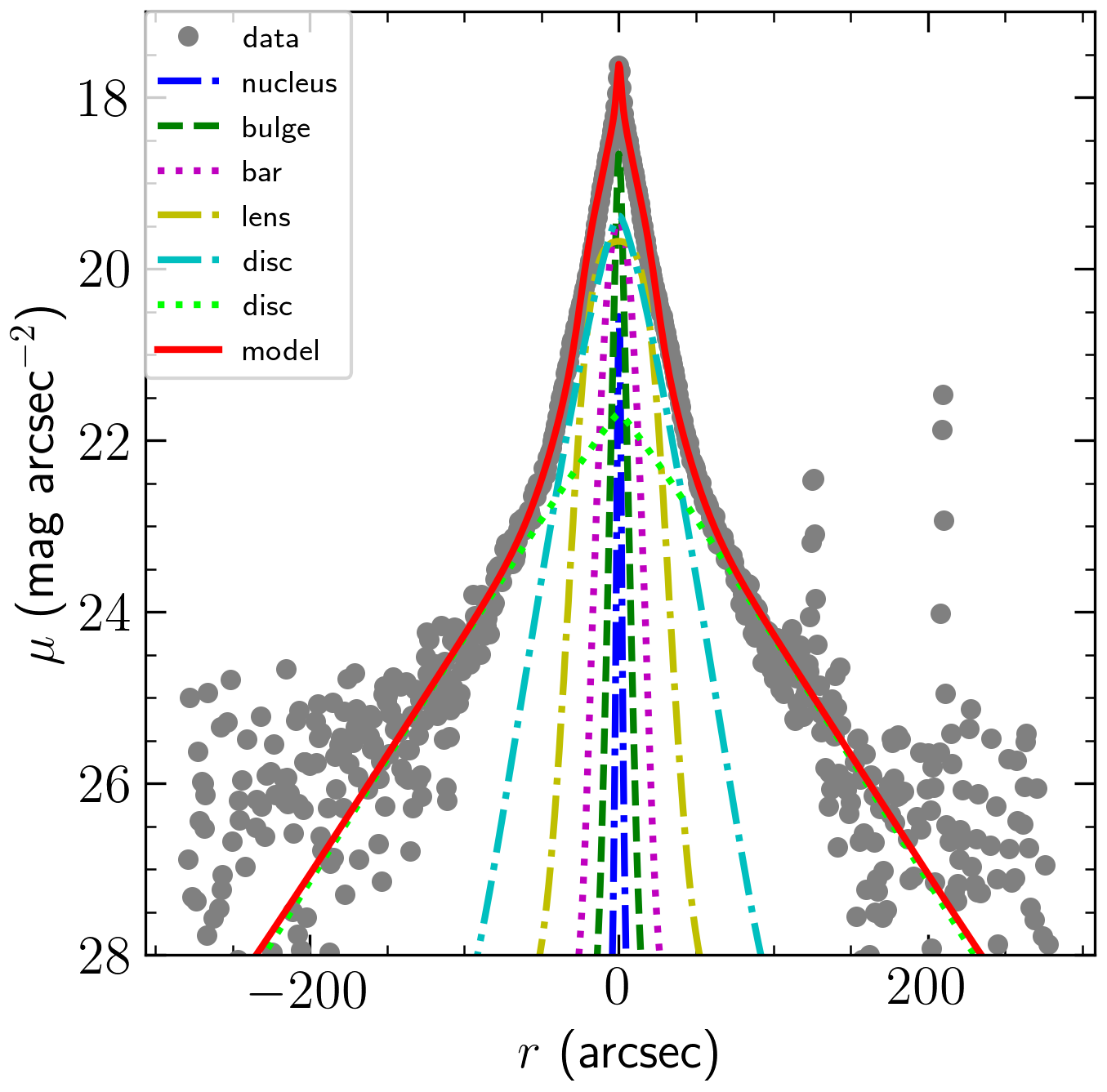}
\includegraphics[height=7cm]{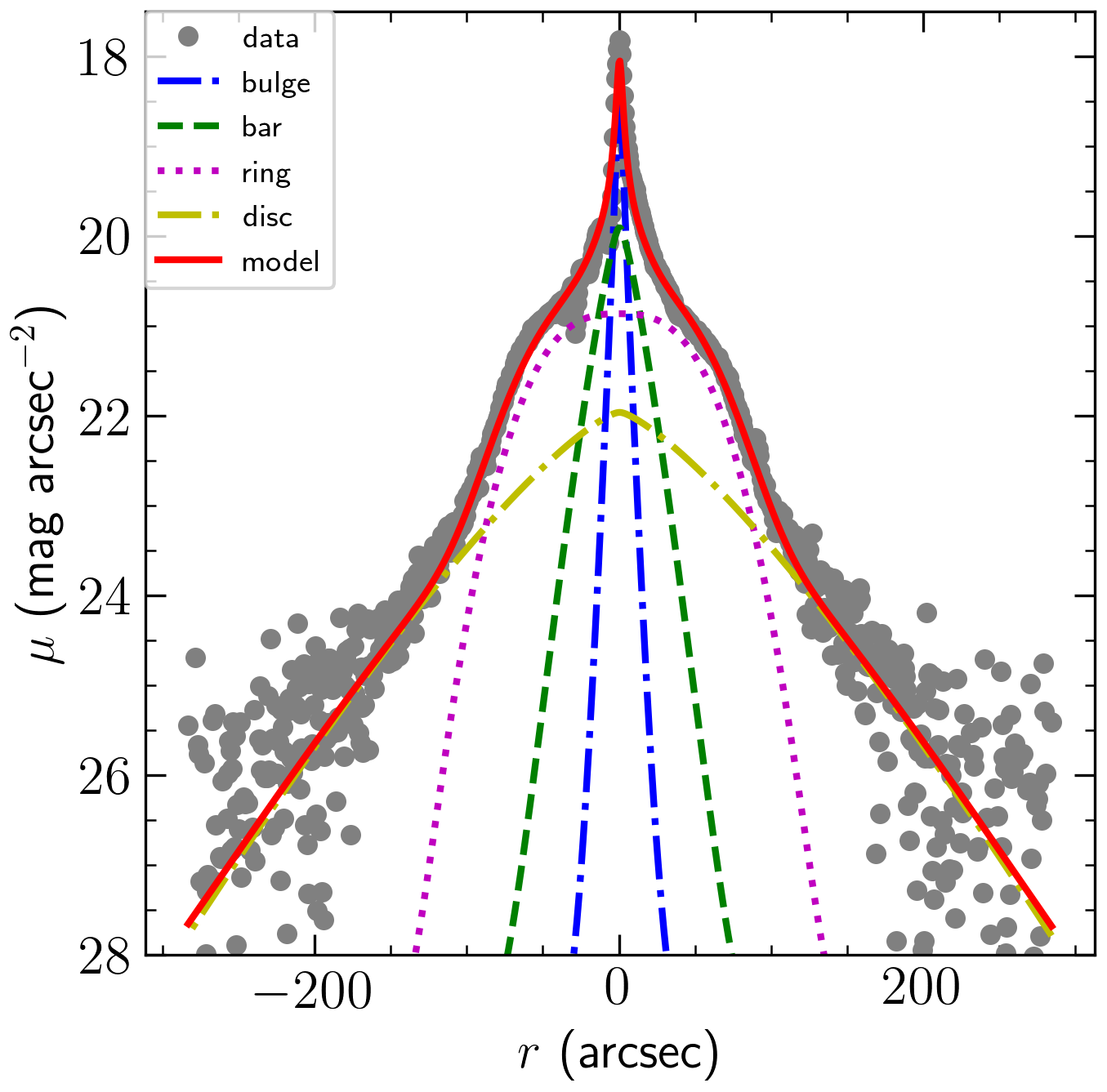}
\caption{Photometric cuts along the major axes of the outer discs for NGC\,4452 (left plot) and NGC\,4469 (right plot) with the superimposed corresponding decomposition models.}
\end{figure*}

\section{Characteristics of warps}
\label{Appendix:warps}
\begin{figure*}
\label{fig:warps}
\includegraphics[height=4cm]{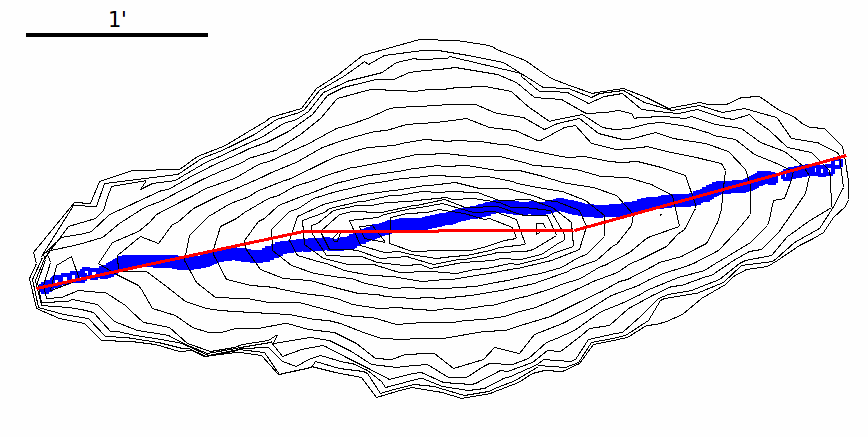}
\includegraphics[height=4cm]{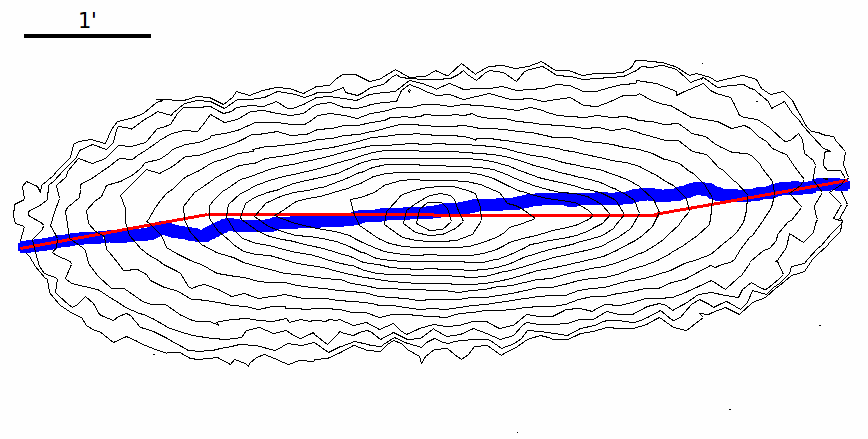}
\caption{Contour maps for NGC\,4452 (left plot) and NGC\,4469 (right plot). The red lines correspond to the best fits of the centre-lines,  based on all isophotes, with piecewise linear functions. The outermost isophote in each plot is of 25.5 mag/arcsec$^2$, with the blue line, which represents the centre-line of this isophote.}
\end{figure*}

\section{Simulations of galaxies with bars}
\label{Appendix:sim}
\begin{figure*}
\label{fig:model_iso}
\centering
\includegraphics[width=8.5cm]{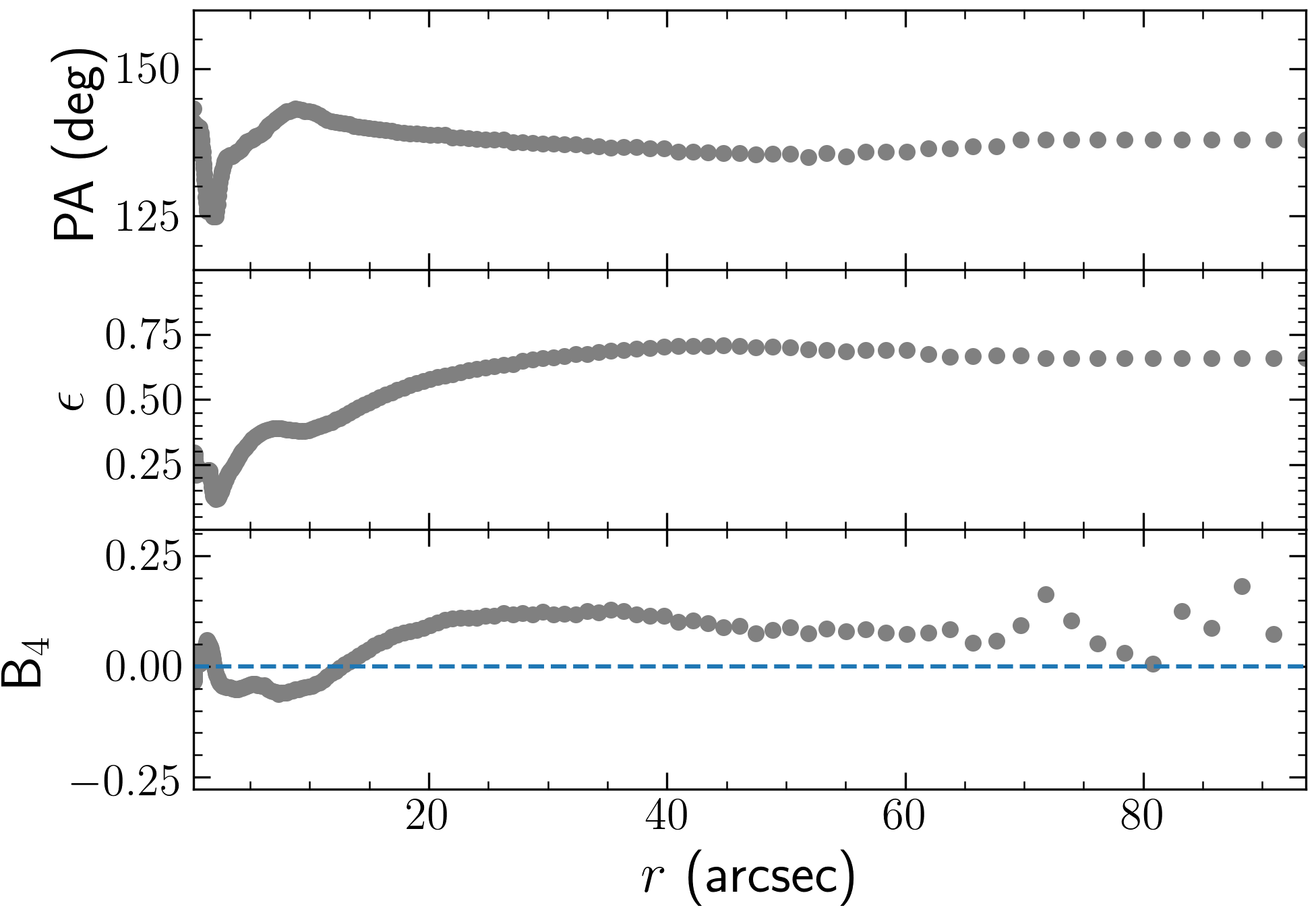}
\includegraphics[width=8.3cm]{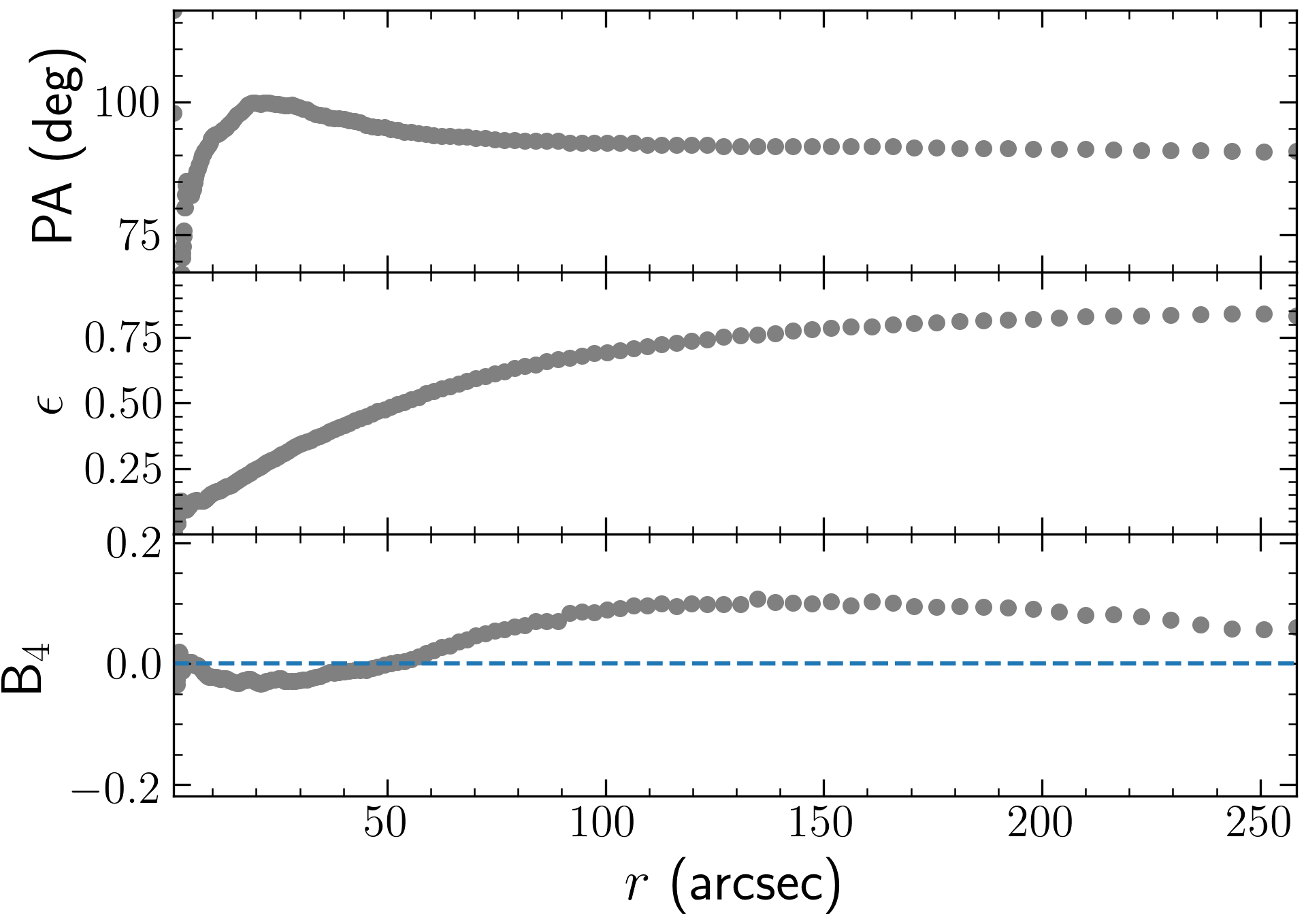}
\caption{Results of the {\tt{IRAF}}/{\tt{ELLIPSE}} fitting for NGC\,3869 (left plot) and the Model~1 (right plot).}
\end{figure*}

Here we briefly describe the details of our simulations which we use in Sect.~\ref{sec:discussion}. To create the initial equilibrium state, we used a script for constructing the equilibrium multicomponent model of a galaxy {\tt{mkgalaxy}} \citep{McMillan_Dehnen2007} from the toolbox for N-body simulation {\tt{NEMO}} \citep{Teuben_1995}. We considered two models; each model consisted of an exponential disc isothermal in the vertical direction,
\begin{equation}
\rho(R,z) = \frac{M_\mathrm{d}}{8\pi h^2 z_0} \cdot \exp(-R/h) \cdot \sech^2(z/(2z_0)) \,,
\label{eq:sigma_disc} 
\end{equation}
 and a dark halo of NWF profile \citep{NFW_article} represented by a set of particles with $N_\mathrm{d}=4 \cdot 10^6$ and $N_\mathrm{h}=4.5 \cdot 10^6$, respectively. The radial velocity dispersion profile of the disc was chosen to obey an exponential law, $\sigma_R = \sigma_0 \cdot \exp(-R/2 h)$, where the $\sigma_0$ value follows from the condition on the Toomre value parameter at $R=2h$, $Q(2h)=Q_0$. First model (Model~1) considered has a relatively thin, cool (in a dynamical sense) and rather massive initial disc with $z_0=0.025\,h$, $Q_0 = 1.2$ and $M_\mathrm{halo}(R<4h)/M_\mathrm{d} = 1$. The second one (Model~2) has a thicker, hotter and lighter disc with $z_0=0.05\,h$, $Q_0=2.0$ and $M_\mathrm{halo}(R<4h)/M_\mathrm{d} = 1.5$. The simulations were carried out in a self-consistent manner, that is, both the disc and the halo were allowed to evolve under the influence of their mutual gravitational field. The equations of motions were solved by the fast numerical integrator {\tt{gyrfalcON}} \citep{Dehnen2002} for about 6 Gyr with an adaptive time step with the maximal value of about $2\cdot 10^{-3}$ Gyrs. Both models developed a strong bar which have a B/PS bulge appearance if seen edge-on. Model images presented in Fig.~\ref{fig:inner_structure} are obtained by rotations of the bar major axis and the disc plane, and integrating the density of the luminous matter along the line of sight. In Fig.~\ref{fig:model_iso} we present the results of the {\tt{IRAF}}/{\tt{ELLIPSE}} fitting for NGC\,3869 and the Model~1 image. We can observe a change of the position angle in the inner region, where the B/PS bulge dominates ($B_4<0$), by approximately $7\degr$ for NGC\,3869 and $10\degr$ for Model~1. Also, note similar behaviours of all light distributions for both galaxies.

\bsp	
\label{lastpage}
\end{document}